\documentclass[numberedappendix,twocolappendix,iop]{emulateapj}
\usepackage{amssymb,amsmath}
\usepackage{color}
\def\bmath#1{\mbox{\boldmath$#1$}}
\def\reviseb#1{#1}
\def\revise#1{#1}
\def\halfwidthfig{1.0}
\def\fullfwidthfig{1.0}

\slugcomment{{\sc Accepted by ApJ:} March 27, 2017} 

\shorttitle{Possible Formation Mechanism for Misaligned Systems} 
\shortauthors{Matsumoto et al.}

\begin{document}

\title{Circumstellar disks and outflows in turbulent molecular cloud cores: possible formation mechanism for misaligned systems}

\author{
  Tomoaki Matsumoto\altaffilmark{1}, 
  Masahiro N. Machida\altaffilmark{2},
  Shu-ichiro Inutsuka\altaffilmark{3},
}

\altaffiltext{1}{Faculty of Sustainability Studies, Hosei University, Fujimi, Chiyoda-ku, Tokyo 102-8160, Japan}
\email{matsu@hosei.ac.jp}
\altaffiltext{2}{Department of Earth and Planetary Sciences, Kyushu University, Fukuoka 812-8581, Japan}
\altaffiltext{3}{Department of Physics, Nagoya University, Chikusa-ku, Nagoya 464-8602, Japan}

\begin{abstract}
We investigate the formation of circumstellar disks and outflows 
subsequent to the collapse of molecular cloud cores with the magnetic field and turbulence.
Numerical simulations are performed by using an adaptive mesh refinement 
to follow the evolution up to $\sim 1000$~yr after the
formation of a protostar.
In the simulations, circumstellar disks are formed around the protostars; those 
in magnetized models are considerably smaller than those in
nonmagnetized models, but their size increases with time.
The models with stronger magnetic field tends to produce smaller disks.
During evolution in the magnetized models, the mass ratios of a disk
to a protostar is approximately constant at $\sim 1-10$\%.
The circumstellar disks are aligned according to their angular momentum, and the 
outflows accelerate along the magnetic field on the $10-100$~au
scale; this produces a disk that is misaligned with the outflow.
The outflows are classified into two types: a magneto-centrifugal wind
and a spiral flow.  In the latter, because of the geometry, the axis of rotation is
misaligned with the magnetic field.
The magnetic field has an internal structure in the cloud cores,
which also causes misalignment between the outflows and the magnetic field on the scale of the 
cloud core.
The distribution of the angular momentum vectors in a core also has a non-monotonic 
internal structure. This should create a time-dependent accretion of 
angular momenta onto the circumstellar disk. 
Therefore, the circumstellar disks are expected to change their orientation 
as well as their sizes in the long-term evolutions.
\end{abstract}


\section{Introduction}

The magnetic field and turbulence play important roles in the early phase
of star formation.
Observations of the magnetic field have indicated that 
molecular clouds and molecular cloud cores
have a large amount of magnetic energy, which is approximately equal to the
kinetic energy \citep{Crutcher99}.
The magnetic field therefore has the potential to control 
the gravitational collapse of cloud cores.
Molecular clouds exhibit broad molecular lines, which are interpreted
as supersonic turbulence \citep{Zuckerman74}.  The turbulence
seems to have a scaling relation such that a smaller scale has a
smaller velocity dispersion \citep{Larson81}.
For high-density cloud cores, 
weak turbulence is suggested by the narrow molecular lines \citep[e.g.,][]{Onishi98}.
As shown by \citet{Burkert00}, such turbulence 
reproduces the observed rotational properties of cloud cores.
This indicates that 
the turbulence contributes angular momentum to the cloud cores, and
it is the origin of the rotation of circumstellar disks and protostars. 

The magnetic field extracts angular momentum from the circumstellar
disks and the infalling envelopes, due to magnetic braking and outflows.
As a consequence, the protostars accrete gas from the disks.  
In the past decade, the existence of the so-called ``magnetic braking catastrophe'' has
been debated  \citep[e.g.,][]{Mellon08}; in this phenomenon, the magnetic field
prevents the formation of the circumstellar disks around the
protostars.  Axisymmetric models have been investigated
intensively, and they show that circumstellar disks are formed around the
protostar. The size of these disks increases with time, but 
they are considerably smaller than nonmagnetized disks \citep[e.g.,][]{Machida11}.
Models in which the magnetic field and
rotation axes are misaligned have also been investigated \citep{Matsumoto04,Li13}.
Few studies have investigated the role of turbulence in disk formation
\reviseb{\citep[e.g.,][]{Seifried12,Seifried13}}.  \citet{Matsumoto11} performed numerical
simulations of the collapse of magnetized turbulent cloud cores, but they
only followed the evolution up to the formation of the first core.
Their simulations reproduced the formation of the outflow but not that of disks,
because the period simulated was too short.

Recent observations have revealed misaligned young stellar objects.
\citet{Hull13} reported that for 
16 Class 0 and Class I sources, the magnetic field in protostellar
cores of $\sim 1000$~au scale is not tightly aligned with outflows.
It has been suggested that in the Class I source L1489 IRS, the
central Keplerian disk is inclined with respect to a flattened
infalling envelope \citep{Brinch07a,Brinch07b}. 
\revise{Moreover, it has been suggested that the class I binary source IRS~43
exhibits a misalignment between the orbit of the binary and the
circumbinary disk \citep{Brinch16}.}
The formation mechanism for such misaligned systems is not yet known.

In this paper, we use high-resolution numerical simulations to investigate the collapse of cloud cores to form
protostars, and we include the effects of
the magnetic field and turbulence of the cloud core.
The formation of circumstellar disks and outflows is also investigated.
In the simulations, it is expected that misaligned protostars will be formed, 
because a priori rotation axes are not assumed in the initial
conditions.
We focus on the early phase of protostar formation, because 
recent observations have provided high-resolution images of very young protostars and circumstellar structures \citep[e.g.,][]{Tokuda14}.

This paper is organized as follows.
Section~\ref{sec:models} presents the model. Section~\ref{sec:methods} discusses the simulation methods, the results are presented in Section~\ref{sec:results}, and they are discussed
in Section~\ref{sec:discussion}.  Our conclusions are presented in
Section~\ref{sec:summary}.

\section{Models}
\label{sec:models}

As the initial state of a molecular cloud core, we consider a
turbulent, spherical cloud threaded by a uniform magnetic field. The
cloud is confined by a uniform ambient gas.  This is similar to the initial state 
considered by \citet{Matsumoto11}. It is further
specified by the initial strength of the turbulence and 
the magnetic field strength, as summarized in Table~\ref{table:model-parameters}.

As a template for a molecular cloud core, we consider a cloud for which the density
profile is that of the critical Bonnor-Ebert sphere \citep{Bonnor1956,Ebert1955}.
We let $\varrho_{\rm BE}(\xi)$ denote the
nondimensional density profile of the critical Bonnor-Ebert sphere
\citep[see][]{Chandrasekhar39}, and then the initial density distribution is
given by
\begin{equation}
\rho(r) = \left\{
\begin{array}{ll}
\rho_0 \varrho_{\rm BE}(r/a) & {\rm for}\; r < R_c\\ \rho_0
\varrho_{\rm BE}(R_c/a) & {\rm for}\; r \geq R_c
\end{array}
\right. \;,
\end{equation}
and
\begin{equation}
a = c_s \left( \frac{f}{4 \pi G \rho_0} \right)^{1/2} \;,
\label{eq:density-enhancement}
\end{equation}
where $r$, $G$, $c_s$, and $\rho_0$ denote the radius, gravitational
constant, isothermal sound speed, and initial central density,
respectively.  The gas temperature is assumed to be 10 K ($c_s =
0.19~\mathrm{km}~\mathrm{s}^{-1}$) .  
The initial central density is set at $\rho_0 = 10^{-18}\,{\rm g}\,{\rm
  cm}^{-3}$, which corresponds to a number density of $n_0 = 2.61
\times 10^5\,{\rm cm}^{-3}$ for an assumed mean molecular weight of
2.3.  
The parameter
$f$ denotes the nondimensional density enhancement factor, and
the critical Bonnor-Ebert sphere is obtained when $f=1$. 
For a given central density, an increase in density by a factor of $f$ is equivalent to 
an enlargement of the spatial scale by a factor of $f^{1/2}$.
We adopt $f=2$ in this paper.
The radius of the cloud is set to be
$R_c = 6.45 a = 0.0434 f^{1/2}\,\mathrm{pc} = 0.0614\,\mathrm{pc}$, 
where the factor 6.45 comes from the
nondimensional radius of the critical Bonnor-Ebert sphere. 
The
density contrast of the initial cloud is $\rho(0)/\rho(R_c) = 14.0$.
The initial freefall timescale at the center of the cloud is
thus $t_{\rm ff} \equiv (3 \pi / 32 G \rho_0)^{1/2} =
6.66\times10^4\,{\rm yr} $.
The mass of the cloud core ($r\le R_c$) is $M_c = 0.89 f^{3/2} M_\sun =
2.51 \,M_\sun$.
The ratio of the thermal energy to the gravitational energy is
estimated as $E_\mathrm{th}/|E_\mathrm{grav}| = 0.84 f^{-1} = 0.42$.
The spherical cloud described above is located at the center of 
the computational domain of 
$x, y, z \in [-2R_c, 2R_c]^3$.

The turbulence is determined by the initial velocity field, and it is not
driven during the course of the simulations; that is, we assume free decay of the turbulence.
The initial
velocity field is incompressible, with a power spectrum of $P(k)
\propto k^{-4}$, and it is generated in accordance with that in \citet{Dubinski95}, where $k$
is the wavenumber.  This power spectrum results in a velocity
dispersion of $\sigma(\lambda) \propto \lambda^{1/2}$, which is in agreement
with the Larson scaling relations \citep{Larson81}, where
$\lambda$ denotes the length scale.  
The root-mean-square (rms) Mach number in the computational domain
is specified by a model parameter, ${\cal M}$;
\begin{equation}
{\cal M} = \left(\frac{1}{V_\mathrm{cd}} \int_{V_\mathrm{cd}} | \bmath{v}^2 | \,dV\right)^{1/2},
\end{equation}
where $V_\mathrm{cd}$ denotes the volume of the computational domain.
We utilized a common velocity field for generating the turbulence in all the models.
We changed the amplitude of the turbulence by changing the Mach number
${\cal M}$.
Note that even if we assume ${\cal M} = 1$, 
the rms Mach number on a 0.06~pc scale (the cloud core scale) is
estimated to be ${\cal M}_c = {\cal M} (R_c/2R_c)^{1/2} = 0.71$, according to the scaling relations, and therefore
the turbulence is subsonic on the cloud core scale.
When ${\cal M} = 0.5$,  the rms Mach number is estimated to be ${\cal M}_c = 0.35$.
Note that weak
turbulence was suggested by the narrow molecular lines  in the dense
cores in Taurus \citep{Onishi98}.
The rms Mach number on the cloud core scale ${\cal M}_c$ for each model is listed in Table~\ref{table:model-parameters}.

The initial magnetic field is uniform in the $z$-direction.
The field strength is given by $B_z = \alpha B_\mathrm{cr} $,
 where $\alpha$ denotes the nondimensional flux-to-mass ratio (see Table~\ref{table:model-parameters}), 
and $B_\mathrm{cr}$
denotes the critical field
strength for the center of the cloud core,
given by $B_\mathrm{cr} = 2 \pi G^{1/2} \Sigma_0$
\citep{Nakano78,Tomisaka88}.
The central column density $\Sigma_0$ is calculated by $\Sigma_0 =
\int_{-R_c}^{R_c} \rho dz = 5.38 \rho_0 a$, where the integral is performed along
a line passing through the center of the cloud core.
In this paper, we consider a magnetically supercritical core ($\alpha < 1$).
The initial field strength is 
estimated to be $B_z = 181 \alpha  f^{1/2} \,\mu\mathrm{G} = 256 \alpha\,\mu\mathrm{G}$; note this uses the model
parameters $\alpha$ and $f$.
Note that the model parameter $\alpha$ is inversely proportional to the 
dimensionless mass-to-flux ratio $\mu$, which is defined as
$\mu = (M_c/\Phi)/(M_c/\Phi)_\mathrm{cr}$; 
its value for each model is listed in Table~\ref{table:model-parameters}.
The magnetic flux is defined by $\Phi = \pi R_c^2 B_z $,
and the mass of the cloud core is defined by $M_c =  4\pi \int _0^{R_c}
\rho r^2 dr$. The critical value is $(M_c/\Phi)_\mathrm{cr} = (2\pi G^{1/2})^{-1}$.
The parameter $\mu$ is the mass-to-flux ratio in the entire
cloud core, while the parameter $\alpha$ reflects the mass-to-flux
ratio only at the central axis.

The barotropic equation of state
is assumed as $P(\rho) = c_s^2 \rho + \kappa \rho ^{7/5}$, where $\kappa
= c_s^2 \rho_\mathrm{cr}^{-2/5}$; the critical density is set at
$\rho_\mathrm{cr} = 10^{-13}\,\mathrm{g}\,\mathrm{cm}^{-3}$
(the corresponding number density is $n_\mathrm{cr} = 2.62\times
10^{10}\,\mathrm{cm}^{-3}$), which is taken from the numerical results
of \citet{Masunaga98}.  

The ohmic dissipation is also considered.
The resistivity $\eta$ is quantitatively estimated according to 
equations~(9) and (10) of \citet{Machida07}.
The magnetic Reynolds number is estimated as
$Re_m = v_f \lambda_J \eta^{-1}$, where
$v_f = [(4/3)\pi G \lambda_J^2\rho]^{1/2}$ is the free-fall velocity,
$\lambda_J = (\pi c_p^2 / G \rho)^{1/2}$ is the Jeans length, 
\revise{and $c_p = (dP/d\rho)^{1/2}$ is the sound speed for the barotropic equation of state.}
The magnetic Reynolds number is less than unity ($Re_m < 1$)
in the region where $n \gtrsim 2\times 10^{12}\,\mathrm{cm}^{-3}$, and thus
the magnetic field is dissipative in that region.
This indicates that the inner portion of the first core is
magnetically dissipative.

We allowed the models listed in
Table~\ref{table:model-parameters} to develop until $t_p \simeq
10^3$~yr for the  magnetized models ($\alpha \ne 0$) and $t_p \simeq 10^4$~yr for
the nonmagnetized models ($\alpha = 0$), where $t_p$ denotes the elapsed time following
formation of a sink particle (a model of a protostar).
\revise{
  The short elapsed times in the magnetized models are due to 
  short time steps of the simulations, which are caused by 
  an extremely fast Alfv\'en speed around the sink particles.  In this study, we
  therefore focus on the early evaluational stages of low-mass star
  formation.
  The recent observational studies have investigated young stars in 
  such early evolutionary stages, e.g., the first hydrostatic core
  candidates \citep[e.g.,][]{Belloche06,Pineda11}.
}

\begin{deluxetable}{lllll}
\tablecaption{Model parameters \label{table:model-parameters}}
\tablehead{
\colhead{Model} &
\colhead{${\cal M}$} &
\colhead{$\alpha$} &
\colhead{${\cal M}_c$} &
\colhead{$\mu$} 
\\
\colhead{} &
\colhead{} &
\colhead{} &
\colhead{} &
\colhead{}
}
\startdata
M05B0   & 0.5 & 0.0  & 0.35 & $\infty$ \\
M1B0    & 1.0 & 0.0  & 0.71 & $\infty$ \\
M05B01  & 0.5 & 0.1  & 0.35 & 2.81 \\
M1B01   & 1.0 & 0.1  & 0.71 & 2.81 \\
M1B025  & 1.0 & 0.25 & 0.35 & 1.12 \\
M05B025 & 0.5 & 0.25 & 0.71 & 1.12 
\enddata
\end{deluxetable}

\section{Methods}
\label{sec:methods}

Gravitational collapse of the cloud cores was calculated using
the three-dimensional adaptive mesh refinement (AMR) code 
SFUMATO \citep{Matsumoto07}.  The
magnetohydrodynamics (MHD) scheme has second-order accuracy in space and time. 
The computational domain was resolved on a base grid of $l =0$ with 
$256^3$ cells.  The maximum grid level was set at $l_\mathrm{max} = 9$.
The cell width was $\Delta x_\mathrm{min} = 0.386$~au on the finest grid, $l = 9$, 
and $\Delta x_\mathrm{max} = 198$~au on the base grid of $l =0$.
The Jeans condition was employed as a refinement criterion:
blocks were refined when the Jeans length was shorter than eight times the cell width, i.e.,
$\lambda_J < 8 \Delta x $, where $\lambda_J$ is the Jeans length.
\revise{ 
  This condition is twice as strict as that originally proposed by
  \citet{Truelove97}, and tested in \citet{Matsumoto07}.  An even more
  strict refinement criterion was proposed by \citet{Federrath11} to
  capture dynamo amplification of the very weak magnetic field in the
  gravitational collapse of primordial clouds. Although it should be
  better to use such a criterion to describe the small scale dynamo
  action in very high beta plasma, a very weak turbulent magnetic
  field does not appear anywhere in the present models. 
}

For the MHD scheme, we adopted the HLLD Riemann solver \citep{Miyoshi05},
though the Roe-type MHD Riemann solver \citep{Fukuda99} was implemented
on the original SFUMATO \citep{Matsumoto07}. The HLLD Riemann solver
is known to be more robust than the Roe-type scheme, because it preserves
positivity.  We did not apply any density floors, even
when the plasma beta became low in the low-density regions. 

For the sub-grid model of a protostar, we used sink particles. 
The details of the implementation of the sink particles is shown in \citet{Matsumoto15a}.
The critical density for sink particle formation is set at 
$\rho_\mathrm{sink} = 1\times 10^{-10}\,\mathrm{g}\,\mathrm{cm}^{-3}$
($n_\mathrm{sink} = 2.62\times 10^{13}\,\mathrm{cm}^{-3}$), 
and the sink radius is set at
$r_\mathrm{sink} = 4 \Delta x_\mathrm{min} = 1.55~\mathrm{au}$.
\revise{
  The minimum cell width $\Delta x_\mathrm{min}$ is determined 
  so that the Jeans length of $\rho_\mathrm{sink}$ is resolved by 
  the cells width $\Delta x_\mathrm{min}$ according to 
  the Jeans condition.
  }

The ohmic dissipation was calculated in accordance with \citet{Matsumoto11b}.  The induction equation for the magnetic field was
split into a hyperbolic term and a parabolic term. The former
corresponds to the ideal MHD and was solved explicitly; the latter corresponds to the ohmic dissipation and was solved
implicitly with the multigrid AMR. This provides second-order spatial accuracy and first-order temporal accuracy.

\section{Results}
\label{sec:results}
\subsection{Overall structures}

\begin{figure*}
\epsscale{\fullfwidthfig}
\plotone{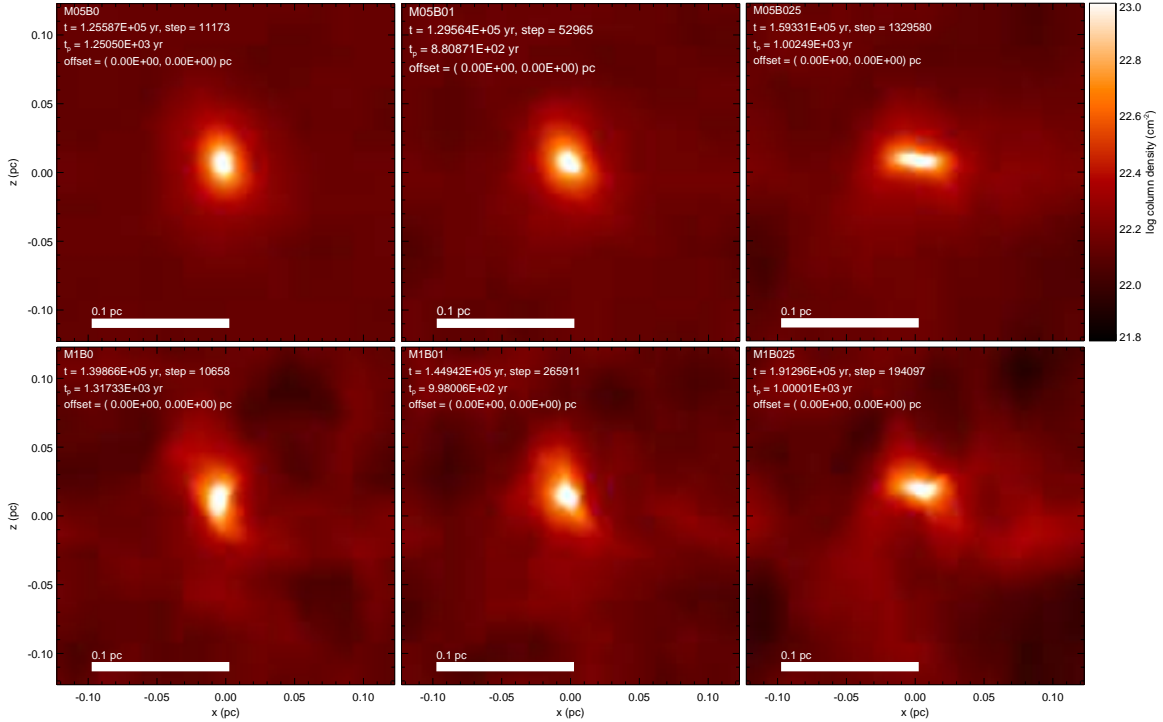}
\caption{
Column density distributions along the $y$-direction for all the models at
$t_p \simeq 1,000$~yr for the entire computational domain.  
The left, middle, and right panels show the models with $\alpha=0.0, 0.1,$ 
and 0.25, respectively.  The top and bottom panels show the models
with ${\cal M} = 0.5$ and 1.0, respectively.
\label{ug.eps}
}
\end{figure*}

\begin{figure*}
\epsscale{\fullfwidthfig}
\plotone{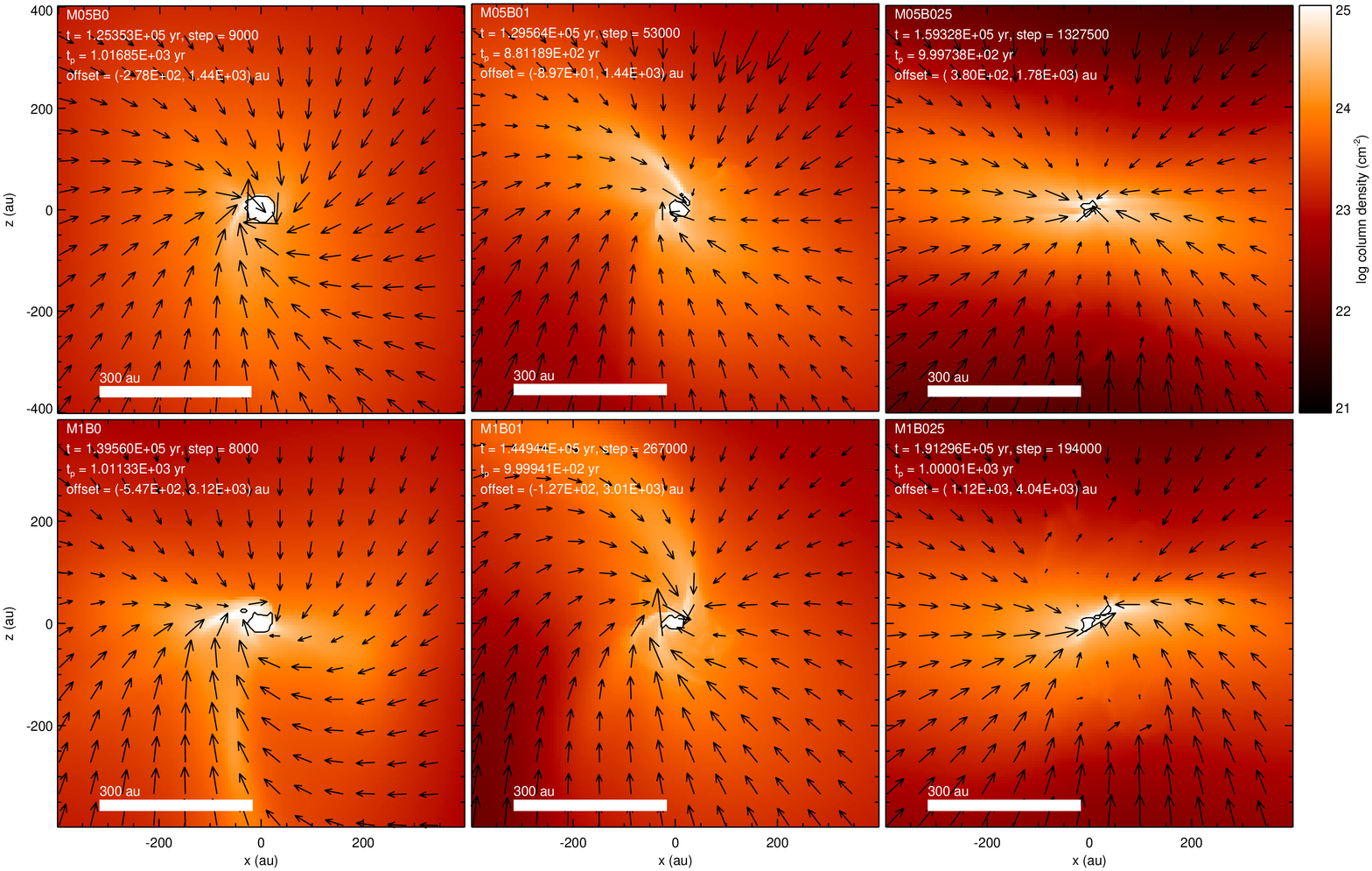}
\caption{
Column density distributions along the $y$-direction 
on the infalling envelope scale
for each of the models at
$t_p \simeq 1,000$~yr.  
The left, middle, and right panels show the models with $\alpha=0.0, 0.1,$ 
and 0.25, respectively.  The top and bottom panels show the models
with ${\cal M} = 0.5$ and 1.0, respectively.
The color scales depict the column density of
$(800~\mathrm{au})^3$ cubes, which  contain the sink particles
at the center.  The arrows denote the density-weighted velocity
distribution. The black contours outline volumes in which $\rho \ge \rho_\mathrm{cr}$.
\label{400au.eps}
}
\end{figure*}

\begin{figure*}
\epsscale{\fullfwidthfig}
\plotone{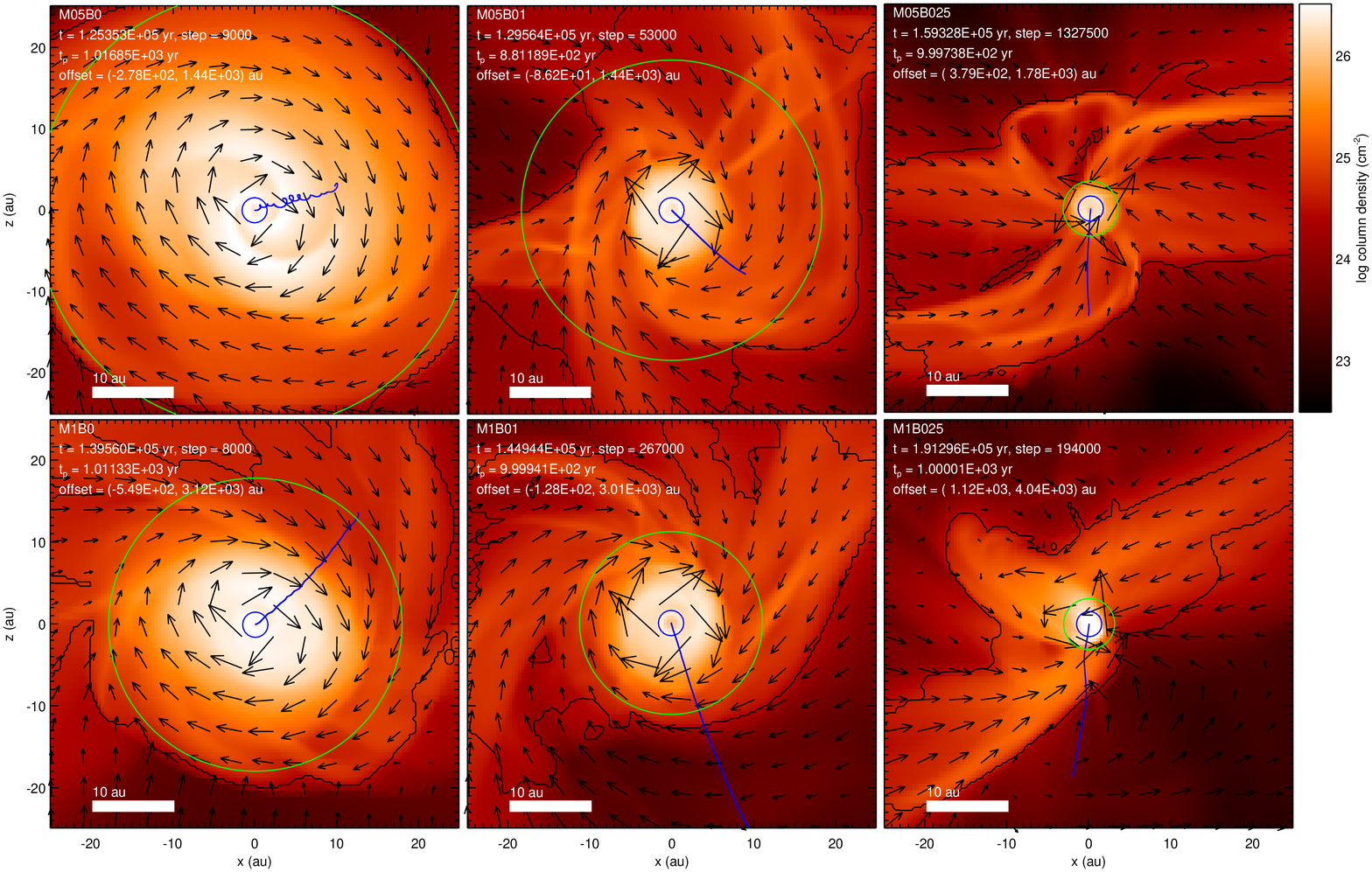}
\caption{
Column density distributions along the $y$-direction 
on the circumstellar disk scale
for each of the models at
$t_p \simeq 1,000$~yr.  
The left, middle, and right panels show the models with $\alpha=0.0, 0.1,$ 
and 0.25, respectively.  The top and bottom panels show the models
with ${\cal M} = 0.5$ and 1.0, respectively.
The color scales depict the column density of
$(50~\mathrm{au})^3$ cubes, which  contain the sink particles
at the center.  The arrows denote the density-weighted velocity
distribution. The black contours outline volumes in which $\rho \ge \rho_\mathrm{cr}$.
The blue circles and the associated blue curves denote the sink radii and 
loci of the sink particles, respectively. 
The green circles are shown with the measured disk radii.
The coordinates are offset so that the center of each panel coincides
with the location of the sink particles. 
\label{25au.eps}
}
\end{figure*}

Figures~\ref{ug.eps}--\ref{25au.eps} show snapshots
of the models at $t_p \simeq 10^3$~yr on the cloud core scale,
the infalling envelope scale, and the circumstellar disk scale, respectively.
Hereafter, we will refer to a circumstellar disk simply as a disk. 
\revise{
The cloud cores collapse on the timescale of $\sim 10^5$~yr.  
In general, a stronger initial turbulence and/or a stronger initial
magnetic field retards the formation epoch of the sink particle. 
The sink particle forms at $t = 1.2\times 10^5$~yr for model M05B0
(the earliest formation)
and $t = 1.9\times 10^5$~yr for model M1B025 (the latest formation). 
}

On the cloud core scale (Figure~\ref{ug.eps}), the cloud cores are deformed
from a spherical shape.
The nonmagnetized models and the weak magnetic field models (the left and middle
panels of Figure~\ref{ug.eps}) produce complex shapes for the cloud
cores because of the turbulence, while the strong magnetic field
models (the right panels of Figure~\ref{ug.eps}) produce flattened cloud cores, which are oriented 
perpendicular to the mean magnetic field (the $z$-direction).
The upper and lower panels in Figure~\ref{ug.eps} show the models with
weak turbulence (${\cal M}=0.5$) and moderate turbulence (${\cal M}=1$), respectively.
Models with moderate
turbulence result in shapes that are more disturbed than those resulting from models with weak turbulence; this is consistent with what was reported
by \citet{Matsumoto11}.

Figure~\ref{400au.eps} shows the column density and velocity
distributions on the envelope scale of $400 ~\mathrm{au}$.  The
nonmagnetized models and weak magnetic field models (left and middle
panels of Figure~\ref{400au.eps}) result in rotating infalling envelopes. The rotation velocities are comparable to the infall
velocities. The angular momentum of the infalling envelopes comes from
the turbulent velocities imposed on the initial conditions.  The
density distributions in the envelopes show spiral structures, and these are caused by the inhomogeneity of the rotation velocities.
On the other hand, the strong magnetic field models with $\alpha =
0.25$ (the right panels of Figure~\ref{400au.eps}) 
produce flattened infalling envelopes that are perpendicular to the
direction of the mean magnetic field.  The arrows in these models 
reflect the silhouettes of the bipolar outflows along the $z$-axis.
Outflow formation is discussed in Section~\ref{sec:Outflow_formation}.

Figure~\ref{25au.eps} shows the column density and velocity
distribution for the central $(50~\mathrm{au})^3$ cubes. 
The disks show almost face-on views for all the models because a common seed was adopted
for the initial turbulence, which provides angular momentum to the
cloud cores.
In each panel, the measured size of the disk is indicated by the green circle; the following two criteria were used to define a disk:
(1)
the density is higher than $\rho_\mathrm{cr}$, and (2) the
rotation speed is considerably faster than the infall velocity. 
The measurement of the disk radii is shown in 
Appendix~\ref{sec:Measurements_of_the_Disk_Radii}.

The size of each disk is correlated with the strength of the magnetic field;
in models with a stronger magnetic field, the disks are smaller, as
shown in Figure~\ref{25au.eps}.
The strong magnetic field model M05B025 has a small disk with a radius of
$R_d = 3.36$~au at $t_p = 10^3$~yr, though this radius is larger than the
sink radius, $r_\mathrm{sink} = 1.55~\mathrm{au}$.
The disk is oriented nearly perpendicular to the $y$-axis
(face-on view in Figure~\ref{25au.eps}), and it is also perpendicular to the
flattened infalling envelopes.  
In the disk, the rotation velocity is considerably greater than the infall velocity.

Model M1B025 has also a small disk, with a radius of $3.08$~au at 
$t_p = 10^3$~yr; note that it is larger than the sink radius.
The disk is elongated toward the upper left in Figure~\ref{25au.eps}.  
The high-density portion of the disk, where $\rho \geq 
10^6 \rho_0  = 10^{-12}\,{\rm g}\,{\rm cm}^{-3}$, is approximately 
axisymmetric, while the edge of the disk with $\rho \sim
10^{-13}\,{\rm g}\,{\rm cm}^{-3}$ has a cometary shape. 

For the weak field models with $\alpha = 0.1$ (models M05B01 and M1B01), the
disks have the face-on views shown in Figure~\ref{25au.eps}.  Each disk is
divided into inner and outer parts.  The inner disk has an approximately
axisymmetric shape with a high density, and its radius
is estimated to be $R_d \sim 5~\mathrm{au}$ at $t_p \sim 10^3$~yr, which
is considerably larger than the sink radius.
The outer disk has spiral arms that wind
around the disk; this occurs because the rotation velocity is greater than the
infall velocity.  
\revise{
  We found that the spiral feature is also caused by the inhomogeneity
  of the ratio between the thermal pressure and magnetic pressure, and
  the inhomogeneity of the rotation velocity along the azimuthal direction.
  The outer disk exhibits the $Q$-value \citep{Toomre64}
  larger than $\sim 2$, indicating that
  it is gravitationally stable.
  }
The Atacama Large Millimeter/submillimeter Array (ALMA) recently observed a circumstellar 
disk with a similar dual structure in ELias~2-27, a Class II star \citep{Perez16}.
\revise{
  The spiral arms of this object is probably caused by the
  gravitational instability \citep[see also][]{Tomida17}, in contrast
  to the models here.
  }

The nonmagnetized models, M05B0 and M1B0, have disks with a large radius of
$R_d \sim (20-30)$~au at $t_p \sim 10^3$~yr. 
For model M1B0, the disk fragments three times:
at $t_p = 7.1\times 10^3~\mathrm{yr}$, $7.8\times 10^3~\mathrm{yr}$, and 
$8.6\times 10^3~\mathrm{yr}$. Two of these fragments merge at
$t_p = 8.9\times 10^3~\mathrm{yr}$, and 
at the end of the calculation period, $t_p \simeq 10^4~\mathrm{yr}$,  there are still three sink particles.
Similar fragmentation is also seen when a filamentary cloud is
used as the initial condition \citep{Matsumoto15b}.
For model M05B0, the disk has not fragmented by the end of the
calculation period, $t_p \simeq 10^4~\mathrm{yr}$. 

\revise{ For all the models, the thicknesses of the disks are resolved
  by more than four cells.  This is consistent with the fact that a
  self-gravitational disk has the scale hight $H = c_p/(2 \pi \gamma
  G \rho)^{1/2} = \lambda_J/( \sqrt{2 \gamma} \pi ) = 0.19 \lambda_J $
  \citep[e.g.,][]{Larson85}, and the Jeans length adopted here
  requires $2H \gtrsim 3 \Delta x_\mathrm{min}$.  }

\subsection{Disk formation}

\begin{figure*}
\epsscale{\fullfwidthfig}
\plotone{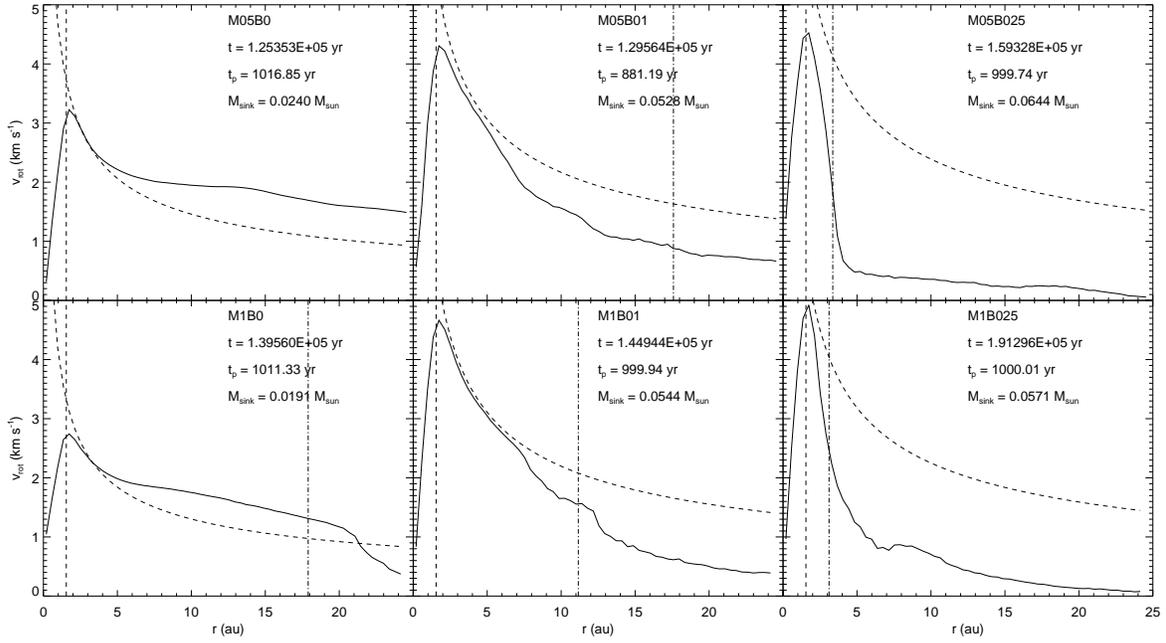}
\caption{
Rotation velocity profiles as a function of the distance from the sink
particles at the stages shown in Figure~\ref{25au.eps}.
The rotation velocities are azimuthally averaged.
The dashed curves show the rotation velocity profiles of the Keplerian
rotation with the masses of the sink particles for comparison.
In each panel, the vertical dashed and dotted-dashed lines indicate the sink radius and 
the measured disk radius, respectively.
\label{rotation_velocity.eps}
}
\end{figure*}

\begin{figure}
\epsscale{\halfwidthfig}
\plotone{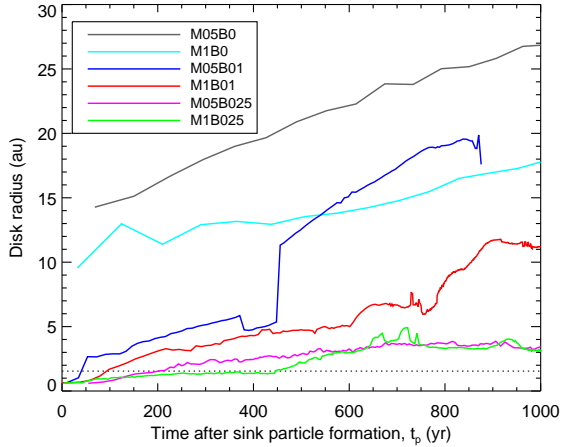}
\caption{
Radius of the circumstellar disks as a function of time following the formation of sink particles.  The horizontal dotted line indicates the sink radius.
\label{disk_radius_plot.eps}
}
\end{figure}

\begin{figure}
\epsscale{\halfwidthfig}
\plotone{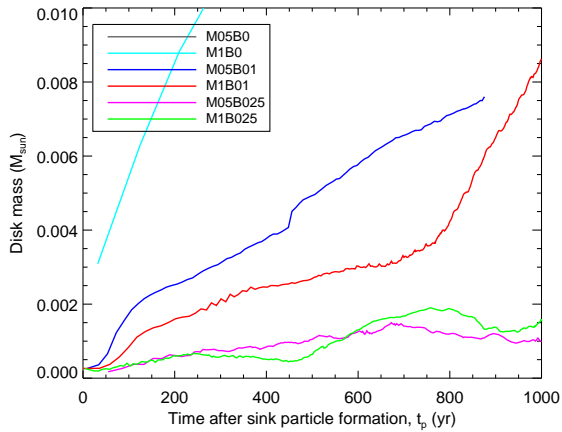}
\caption{
Mass of the circumstellar disks as a function of time following the formation of sink particles.
\label{disk_radius_plot_dmass.eps}
}
\end{figure}

\begin{figure}
\epsscale{\halfwidthfig}
\plotone{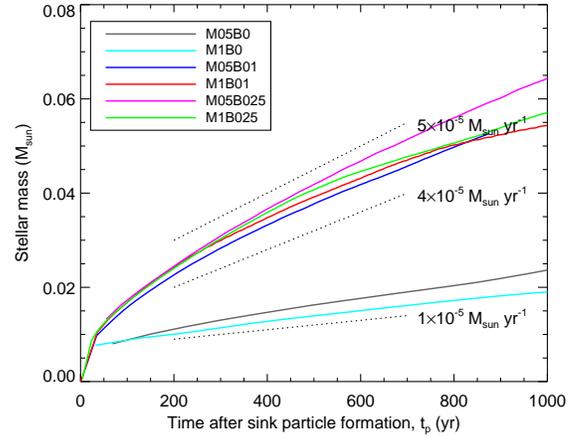}
\caption{
Mass of sink particles as a function of time following the formation of sink particles.  For comparison, the dotted lines indicate the increase in mass due to accretion
rates of $\dot{M} = 5\times 10^{-5}\,M_\sun\,\mathrm{yr}^{-1}$, 
$4\times 10^{-5}\,M_\sun\,\mathrm{yr}^{-1}$, and
$1\times 10^{-5}\,M_\sun\,\mathrm{yr}^{-1}$.
\label{disk_radius_plot_pmass.eps}
}
\end{figure}

\begin{figure}
\epsscale{\halfwidthfig}
\plotone{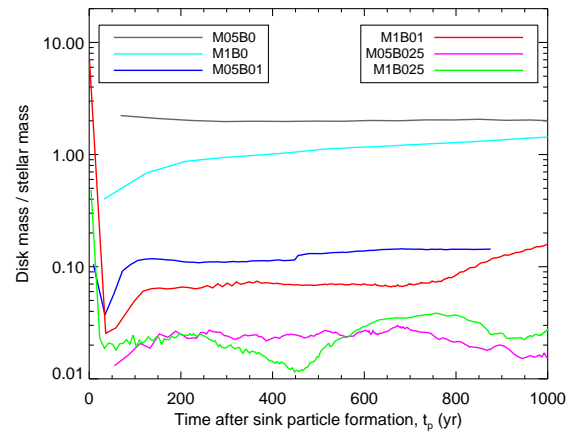}
\caption{
Ratio of the disk mass to the sink particle mass 
as a function of time following the formation of sink particles. 
\label{disk_radius_plot_ratio.eps}
}
\end{figure}

\revise{
Figure~\ref{rotation_velocity.eps} shows the rotation velocity
profile as a function of the distance form the sink particle for
each of the models. The rotation velocity profile is obtained as
follows. The orientation of the disk axis is determined according to
the total angular momentum for a volume of $\rho \ge \rho_\mathrm{cr}$.  
According to the disk orientation, 
the azimuth velocity $v_\varphi$ is calculated
with respect to the sink particle, and 
it is averaged with a density weight along the vertical direction of
the disk for a volume of $\rho \ge \rho_\mathrm{cr}$.
Finally, the density-weighted $v_\varphi$ is azimuthally averaged to
obtain the rotation velocity profile shown in Figure~\ref{rotation_velocity.eps}.
The nonmagnetized models (M05B0 and M1B0) exhibit rotation
velocity profiles faster than those of the Keplerian rotation because of the
massive disks.  The self-gravity of the disk increases the
rotation velocity.
For the weak field models (M05B01 and M1B01), the
inner parts of the disks ($r \lesssim (5 - 7) $~au) exhibit the Keplerian rotation. 
These regions correspond to those showing an approximately
axisymmetric shape in Figure~\ref{25au.eps}.  
In the outer parts of the disks, rotation velocity is slower than that of the
Keplerian rotation, and the infall motion is observed there as shown in Figure~\ref{25au.eps}.
For the strong field models (M05B025 and M1B025), the disk sizes are
as small as $\sim 3$~au, and the rotation velocities are close to
those of the Keplerian rotation near the sink radius. 
We confirmed that the centrifugal force is a dominant force against
the gravity in the disks, and the rotation velocity is considerably
faster than the infall velocity.
Note that, for all the models, the regions within the sink radius exhibit slower
rotation than the Keplerian rotation because of the softening of the
gravity therein. 
}

Figure~\ref{disk_radius_plot.eps} shows the increase in the radius of the
disks as a function of time following the formation of sink particles for each of the models.
The strong magnetized models ($\alpha=0.25$) have disks that are smaller than those for
the nonmagnetized models ($\alpha=0$).
The disk radius for the strong field models remain at $R_d \sim 3$~au 
during the simulation period of $6\times 10^2\,\mathrm{yr} \lesssim t_p \lesssim 10^3\,\mathrm{yr}$.
In the weak magnetic field models ($\alpha=0.1$), the disk radius increases with
considerable undulations,  which are caused by the dynamical changes in
the spiral arms.  The radius of the disks is sensitive to changes in the
spiral arms associated with the outer parts of the disks. 

Figure~\ref{disk_radius_plot_dmass.eps} shows the mass of the
disks as a function of time following the formation of sink particles.  The stronger magnetic field models show slower growth of
the disk mass, and this is roughly independent of the Mach number ${\cal M}$.  
The models with $\alpha=0.25$ have a disk mass of
$M_d \sim 10^{-3}\,M_\sun $  at $t_p = 10^3$~yr, 
while models with $\alpha=0.1$ have a disk mass of $M_d \sim 8 \times 10^{-3}\,M_\sun$.

Figure~\ref{disk_radius_plot_pmass.eps} shows the mass of the sink
particles as a function of time following the formation of sink particles.
The growth of the sink particles exhibits a clear tendency:
magnetic field models have higher accretion rates than do the
nonmagnetized models; all the magnetized models have
accretion rates of approximately $(4-5)\times
10^{-5}\,M_\sun\,\mathrm{yr}^{-1}$, while the 
nonmagnetized models have rates of $\sim 1 \times
10^{-5}\,M_\sun\,\mathrm{yr}^{-1}$. 
\revise{
  Note that we follow only the first $\sim 10^3$~yr of evolution of
  the sink particles, and thus we cannot determine the final stellar masses in our models.
  }

Figure~\ref{disk_radius_plot_ratio.eps} shows the ratio of the disk
mass to the sink particle mass for each model.  The strong magnetic
field models ($\alpha = 0.25$) exhibit a low ratio of $\sim 0.02-0.03$,
indicating that a strong magnetic braking efficiently extracts angular momentum
from the disks.  The weak magnetic field models
($\alpha=0.1$) have a mass ratio of $\sim 0.1$, which is an order of
magnitude larger than that of the strong field models.
The nonmagnetized models ($\alpha=0$) have ratios that are on the order of unity,
and such large ratios induce fragmentation of the disk (model M05B0).


\subsection{Outflow formation}
\label{sec:Outflow_formation}

\begin{figure*}
\epsscale{\fullfwidthfig}
\plotone{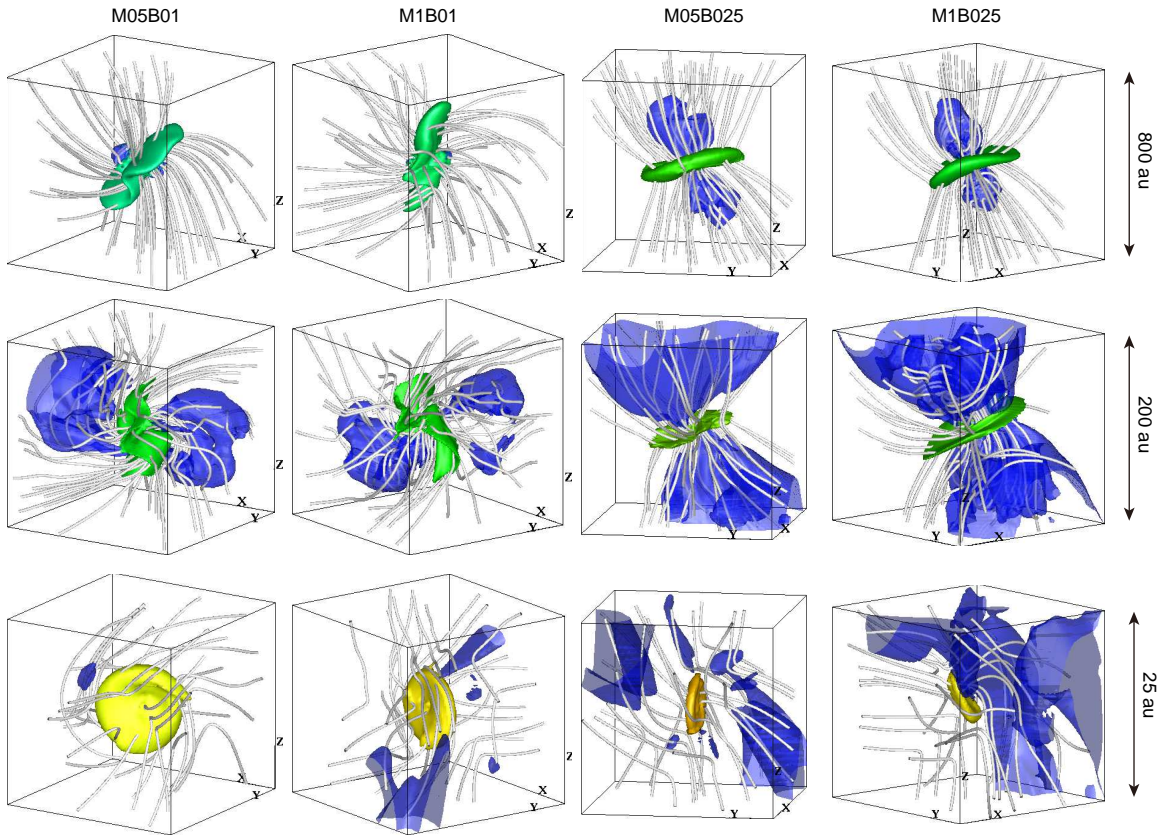}
\caption{
Outflows, magnetic field, and density distribution at $t_p = 700$~yr
for models M05B01, M1B01, M05B025, and M1B025, from left to right.
The upper, middle, and lower panels show the regions of $(800~\mathrm{au})^3$,
$(200~\mathrm{au})^3$, and $(25~\mathrm{au})^3$, respectively.
The blue isosurfaces indicate where the radial velocity is $v_r = 2 c_s$
($v_r = 0.38\,\mathrm{km\,s}^{-1}$) in the upper and middle panels,
and $v_r = 5 c_s$ ($v_r = 0.95\,\mathrm{km\,s}^{-1}$) in the lower panels.
The green isosurfaces indicate where the density is $\log (\rho/\rho_0) = 3$ ($n =
2.6\times10^8\,\mathrm{cm}^{-3}$) in
the top panels, and 4.5 ($n = 8.3\times 10^9\,\mathrm{cm}^{-3}$) in the middle panels; this indicates the infalling envelopes.
The yellow isosurfaces indicate where the density is
$\log (\rho/\rho_0) = 6.5$ ($n = 8.3\times10^{11}\,\mathrm{cm}^{-3}$) in
the lower panels, representing the circumstellar disks. 
The tubes indicate the magnetic field lines. 
\label{jets.eps}
}
\end{figure*}

\begin{figure}
\epsscale{\halfwidthfig}
\plotone{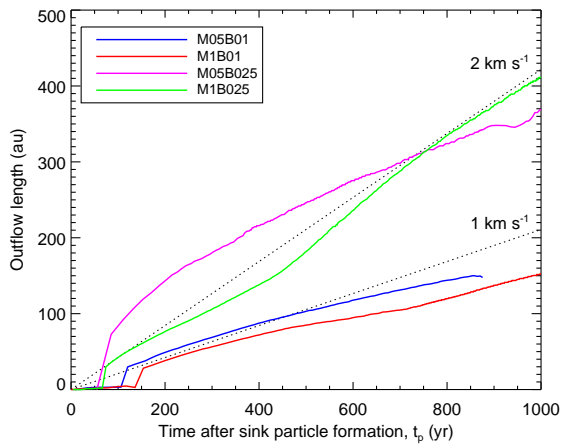}
\caption{
Outflow length as a function of time following the formation of sink particles
($t_p$) for the magnetized models.
For comparison, the dotted black lines indicate lengths extending at speeds of
$1~\mathrm{km\,s}^{-1}$ and $2~\mathrm{km\,s}^{-1}$.
\label{outflow_len_plot.eps}
}
\end{figure}

Figure~\ref{jets.eps} shows outflows at $t_p = 700$~yr for 
the magnetized models.
The strong magnetic field models (M05B025 and M1B025) have outflows
that extend further than do those for the weak magnetic field models (M05B01 and M1B01).
The envelopes for the weak magnetic field models are disturbed by the turbulence,
as shown on the 800~au scale, while the strong magnetic field models have disk-shaped
infalling envelopes.
For all the models, the outflows extend in approximately the
mean direction of the magnetic field on that scale.
The direction of the magnetic field depends on its initial strength, and when the field is strong, the
initial direction is approximately maintained ($z$-direction).  In the weak magnetic field
models, the magnetic field lines become steeply inclined with respect
to the initial direction, and this has been reported by \citet{Matsumoto11}.

The outflows are not always aligned with the 
disks, for which the orientations are shown in the lower panels of
Figure~\ref{jets.eps}.  For models M05B025 and M1B025, the lower panels of
Figure~\ref{jets.eps} are almost edge-on views of the disks,
and the outflows extend nearly vertically in the upper and
middle panels of Figure~\ref{jets.eps}.  This indicates that the
outflows are roughly perpendicular to the disk axes for models M05B025
and M1B025.
In Figure~\ref{jets.eps}, the face-on disk of Model M05B01 is shown in the lower panel, and the 
outflow extends horizontally, as shown in 
the middle panel.
This indicates that there is a misalignment between the disk and the outflow.
On the other hand, model M1B01 produces a disk that is roughly aligned with
the outflow. 
Note that, for each of the models, the infalling envelope on the $\gtrsim 100$~au scale is aligned with the outflow, as shown in
Figure~\ref{jets.eps}.  The envelope is oriented
perpendicular to the magnetic field on that scale, and the outflow
extends along the magnetic field, 
as reported by \citet{Matsumoto04}.

\begin{figure*}
\epsscale{0.5}
\plotone{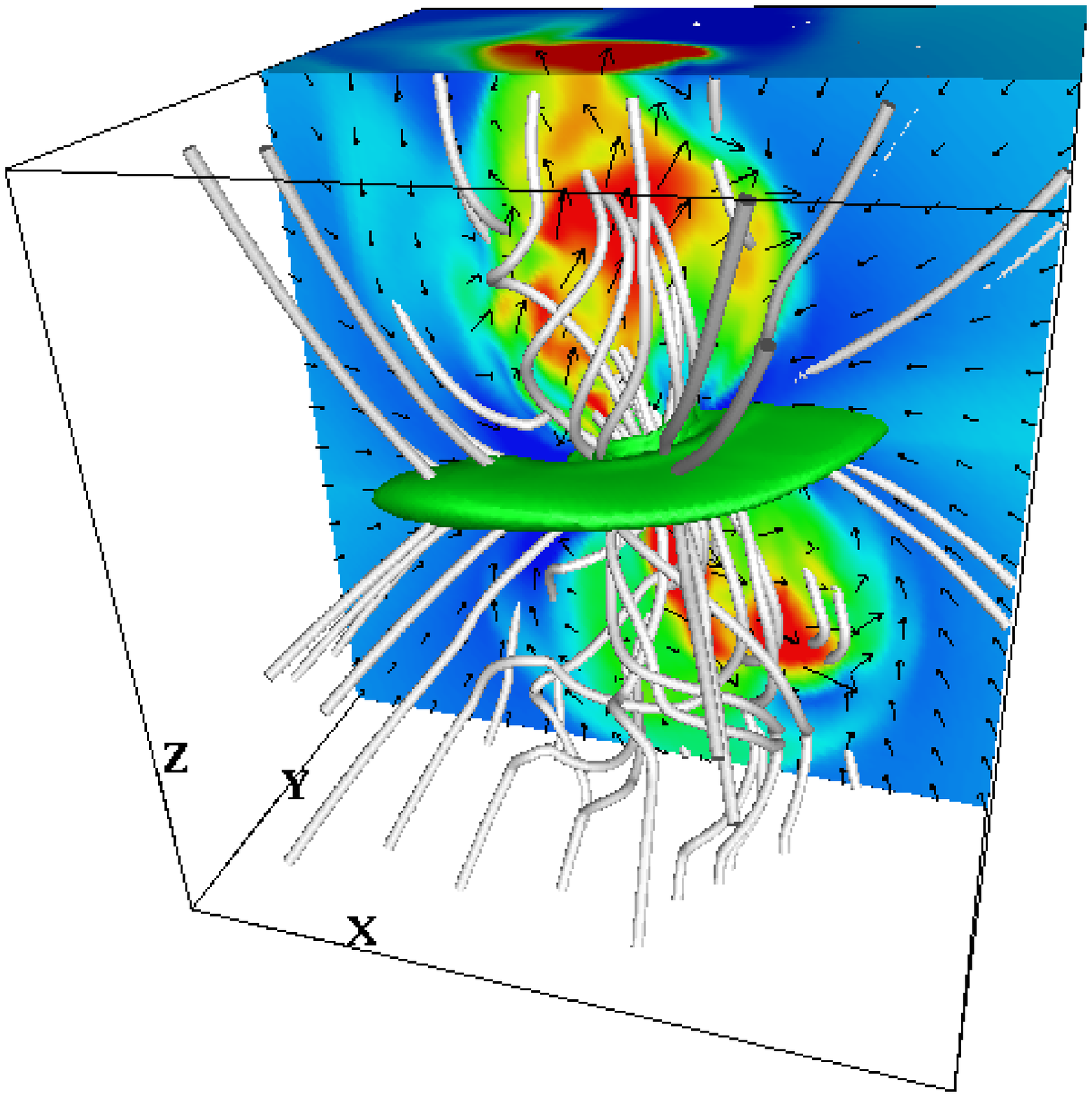}
\plotone{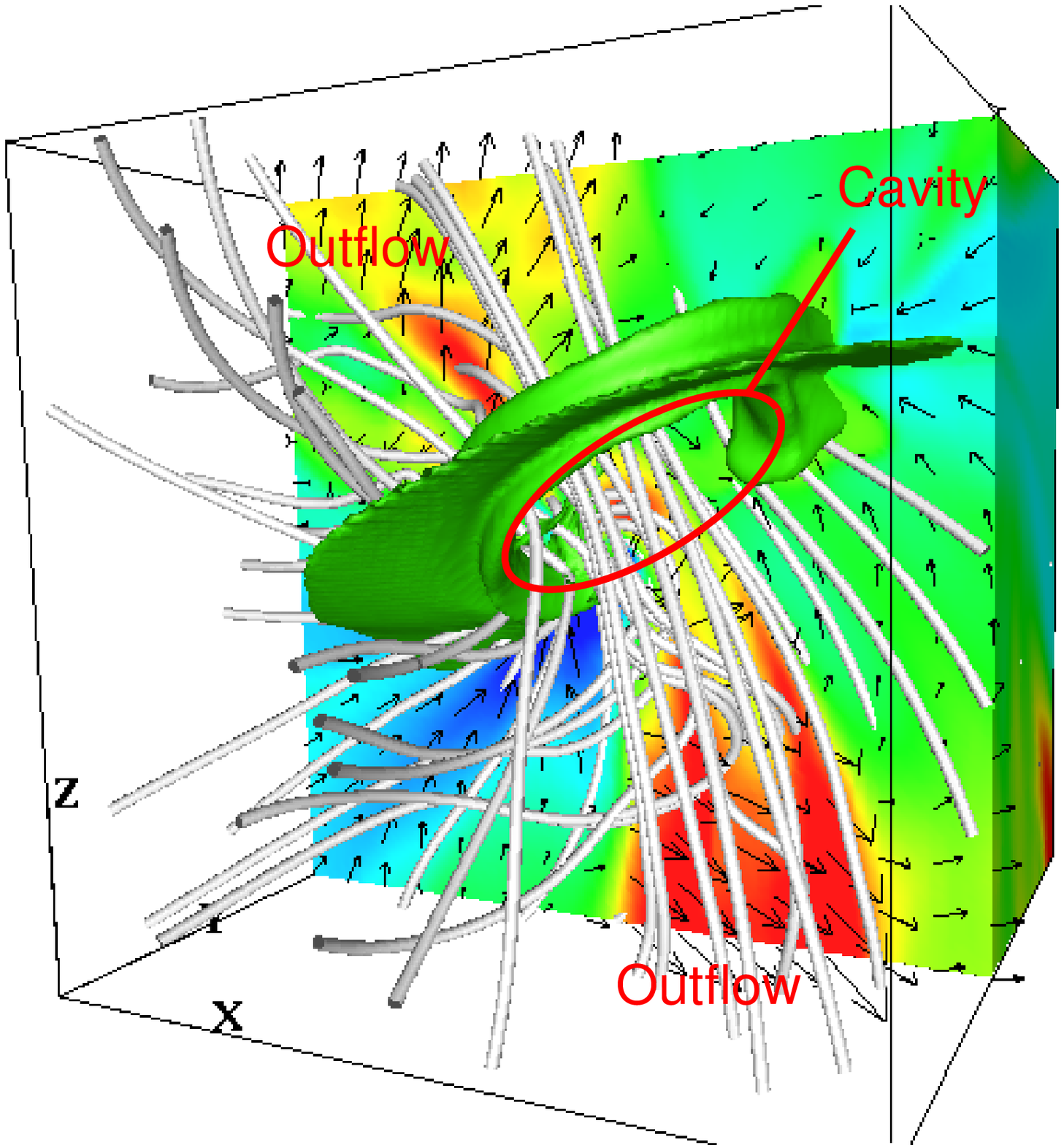}
\caption{
  Outflow and envelope structures for model M1B025 at $t_p = 700$~yr.
  The left and right figures show the regions of
  $(400~\mathrm{au})^3$ and $(100~\mathrm{au})^3$, respectively.
  The green isosurfaces indicate where the density is $\log (\rho/\rho_0) = 3.5$ ($n = 8.9\times10^8\,\mathrm{cm}^{-3}$)
 in the left figure and 4.9 ($n = 2.1\times10^{10}\,\mathrm{cm}^{-3}$)
in the right figure.  The color scale indicates the radial velocity
  distribution on the plane that includes the sink particle.
  The color corresponds to a velocity from $-5c_s$ to $8c_s$ (from $ -0.95\,\mathrm{km\,s}^{-1}$ to $1.5\,\mathrm{km\,s}^{-1}$) in 
  the left figure and from $-10c_s$ to $10c_s$
  (from $ -1.9\,\mathrm{km\,s}^{-1}$ to $1.9\,\mathrm{km\,s}^{-1}$) in 
  the right figure.
  The arrows indicate the velocity distribution on those planes.
\label{jet_vr_M1B025.eps}
 }
\end{figure*}

Figure~\ref{outflow_len_plot.eps} shows the length of the outflow as
a function of time following the formation of sink particles.  The length
is measured from the sink particle to the maximum distance of the outflow region, where the outflow region is
defined by a volume of $v_r \ge 2c_s$.
This plot indicates that the strong magnetic field models result in more rapid
growth than do those with a weak magnetic field.  The growth rates are
$\sim 2\,\mathrm{km\,s}^{-1}$ and $\sim 1\,\mathrm{km\,s}^{-1}$ for
the strong and weak magnetic field models, respectively.
These rates are consistent with typical gas velocities in the
outflow (Figure~\ref{jet_vr_M1B025.eps}).

\subsection{Cavity formation}
\label{sec:Cavity_formation}

\begin{figure}
\epsscale{\halfwidthfig}
\plotone{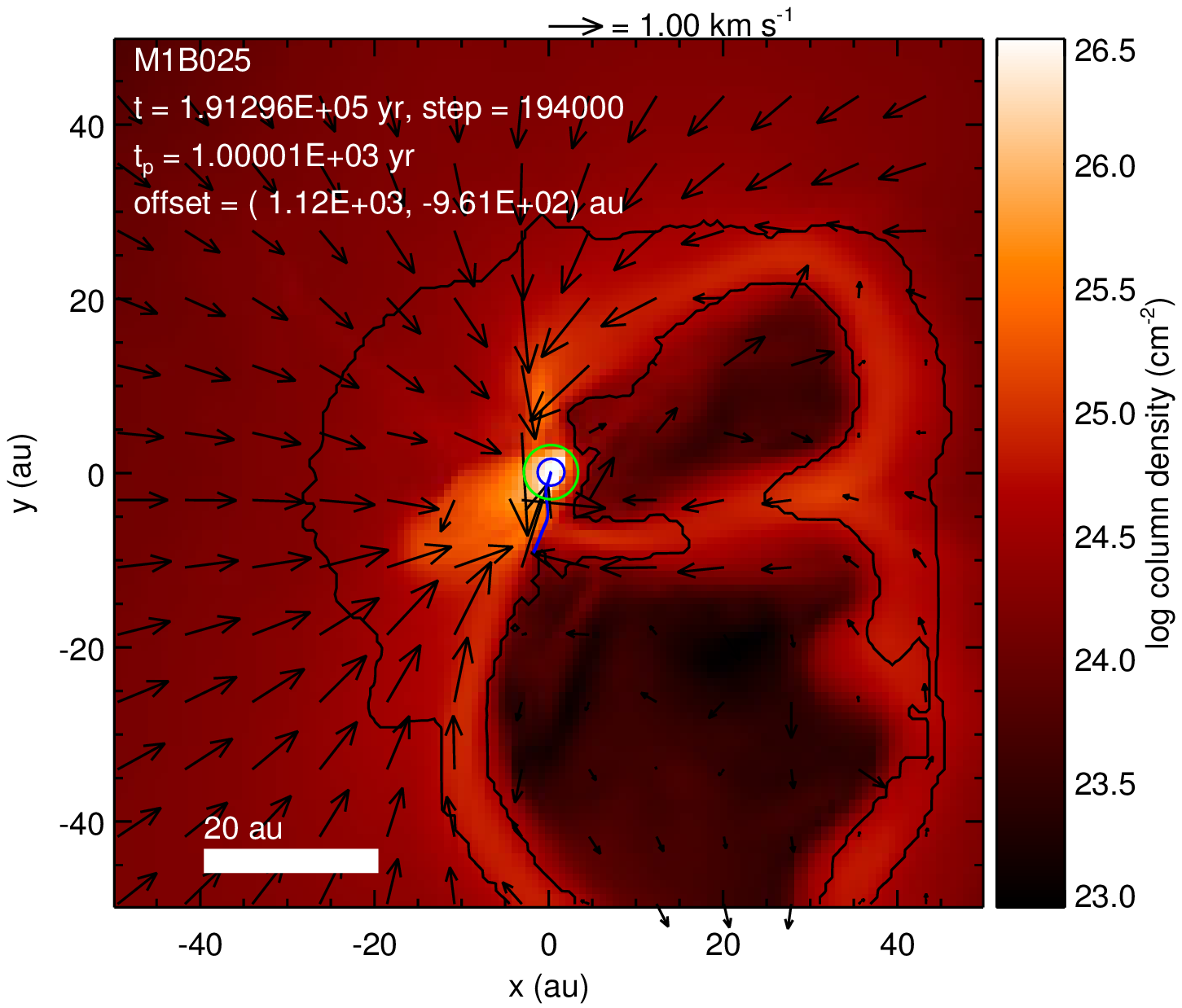}
\caption{
  Column density distribution along the $z$-direction for model M1B025
  at $t_p = 1,000$~yr.
The color scale indicates the column density for
a $(100~\mathrm{au})^3$ cube centered on the sink particle.  The arrows indicate the density-weighted velocity
distribution. The black line surrounds a volume of $\rho \ge \rho_\mathrm{cr}$.
The green circle and the associated blue curve indicate the measured disk radius and 
the locus of the sink particle, respectively. 
\label{colmndensity_M1B025_50au.eps}
 }
\end{figure}

One of the prominent features in the strong field models is a cavity
structure in the envelope. 
Figure~\ref{jet_vr_M1B025.eps} shows the outflow and envelope
structures for model M1B025 on the scales of 400~au and 100~au.
On the 400~au scale, the flattened envelope is perpendicular to the outflow,
which is associated with the helical magnetic field lines. 
On the 100~au scale, the flattened envelope has a cavity in which the magnetic field lines are straight.
The bipolar outflow is not
associated with the straight magnetic field lines of the cavity; instead, it  
is associated with the helical
magnetic field lines, which thread the disk around the sink
particles. 
The cavity is created beside the disk, as shown in Figure~\ref{colmndensity_M1B025_50au.eps}.
The radius of the cavity increases with time, and it increases up to $\sim 50$~au
by $t_p \simeq 10^3$~yr.

The cavity is caused by the magnetic pressure in the following way.
The magnetic pressure is higher inside the cavity than it is in the other
regions of the envelope, and thus it pushes the gas away from
the flattened envelope.  The gas accumulates on the rim of the
cavity, which then has a higher density than the other regions,
as shown in Figure~\ref{colmndensity_M1B025_50au.eps}.
Inside the cavity, the gas moves outward at a velocity of $\sim 0.2 - 0.5\,\mathrm{km\,s}^{-1}$.
The velocity at which the rim extends is typically
$0.2\,\mathrm{km\,s}^{-1}\, (= 50~\mathrm{au}/10^3~\mathrm{yr})$,
which is roughly equal to the speed of sound.

Model M05B025 also results in cavities in the envelope, and the size of
the cavities ($\sim 10$~au) is smaller than that for model M1B025 when compared at
the same time ($t_p = 10^3$~yr).  In the upper right panel of
Figure~\ref{25au.eps}, cavities can be seen 
at both the upper and lower sides of the disk, but their density
contrast is less than that for the cavities seen in model M1B025.
The weak magnetic field models  M05B01 and M1B01 do not produce 
cavities in the envelopes.

Similar cavity formation has been reported in 
recent MHD simulations
\citep[e.g.,][]{Zhao11,Joos12,Krasnopolsky12,Machida14}.
In these simulations, the cavities are formed 
whether or not sink particles are implemented, and
whether or not magnetic diffusion is considered.
Our simulations suggest that the formation of a cavity
depends on the magnetic field strength.
The cavities reproduced in these simulations likely correspond to 
the magnetic wall that has been 
predicted by theoretical studies \citep{Li96,Tassis05}.

\section{Discussion}
\label{sec:discussion}
\subsection{Disk growth}

Our simulations indicate that a cloud core with a stronger magnetic
field produces a disk with a smaller radius.  On the timescale 
of $t_p \sim 10^3$~yr, this is roughly consistent with long-term
simulations of magnetized rotating cloud cores performed by \citet{Machida11}
\citep[see also][]{Machida13}. However, their simulations tend to result in larger disk radii than those produced by the models considered here.
In their simulations, the disk radii exceed 10~au by $t_p = 10^3$ 
when $\mu = 1$, and reach 20~au when
$\mu=3$ \citep[see ][Figure~9a]{Machida11}
Meanwhile, the models here had a disk radius of $3-4$~au when
$\mu = 1.12$ and $10-20$~au when $\mu =
2.81$.  

The difference in the disk radii is due to
the difference in the initial distribution of the angular momentum.
\citet{Machida11} assumed a uniform rotation, which provides a specific
angular momentum distributed as  $j \propto r^2$.
On the other hand, the turbulence produces a velocity distribution of
$\Delta v \propto r^{1/2}$, due to the Larson scaling relations
assumed here \citep{Larson81}, and the distribution of the specific angular momentum
is expected to be $j \propto r^{3/2}$.
The differences seen in the initial angular momentum distributions are
therefore larger for larger radii. This has a greater effect on the disk radius in the
later stages of the accretion phase, 
because the angular momentum of the infalling gas has a strong impact on
disk evolution \citep[e.g.,][]{Vorobyov15}.
This suggests that, for the models here, the disk radius is expected to increase
with time, e.g., $t_p = 10^5$~yr, but it will still be smaller than
that produced by a model with uniform rotation ($\sim 100$~au).  Small disks with a radius of less than 100~au
are expected to be produced by the magnetized turbulent model.

Recent high-resolution observations have revealed Keplerian disks around
Class 0 and Class I protostars \citep[e.g.,][]{Jorgensen09,Tobin12}.
\citet{Ohashi14} recently suggested that the Class 0 protostar L1527 IRS
has a Keplerian disk with a small radius of 54~au. 
A more extended disk has been suggested 
in the Class 1 source TMC 1A, which has an estimated radius of 100~au \citep{Aso15}. 
Disks that have a radius that exceeds 100~au have also been suggested in 
the Class 0 source VLA1623A \citep{Murillo13} and 
the Class I source L1489 IRS \citep{Hogerheijde01,Yen14}.
Such a variety of Keplerian disks may be responsible for the variety of
magnetic fields and turbulence in natal cloud cores.

\subsection{Alignment between an outflow, a disk, and an envelope}

The various models examined here have shown that some disks are misaligned with
the outflow, as described in Section \ref{sec:Outflow_formation}.  
\citet{Matsumoto04} investigated the direction of
outflows when the initial magnetic field is misaligned with the
rotation axis on the cloud core scale \citep[see also][]{Hennebelle09,Ciardi10}. 
Their simulations indicate that the outflows are extended in the
direction parallel to the local magnetic field, even when this direction
is not aligned with the magnetic field on the cloud core scale.
This causes the outflow on the $\sim 100$~au scale to be misaligned with
the magnetic field on the cloud core scale.

\begin{figure*}
  \epsscale{0.5}
  \plotone{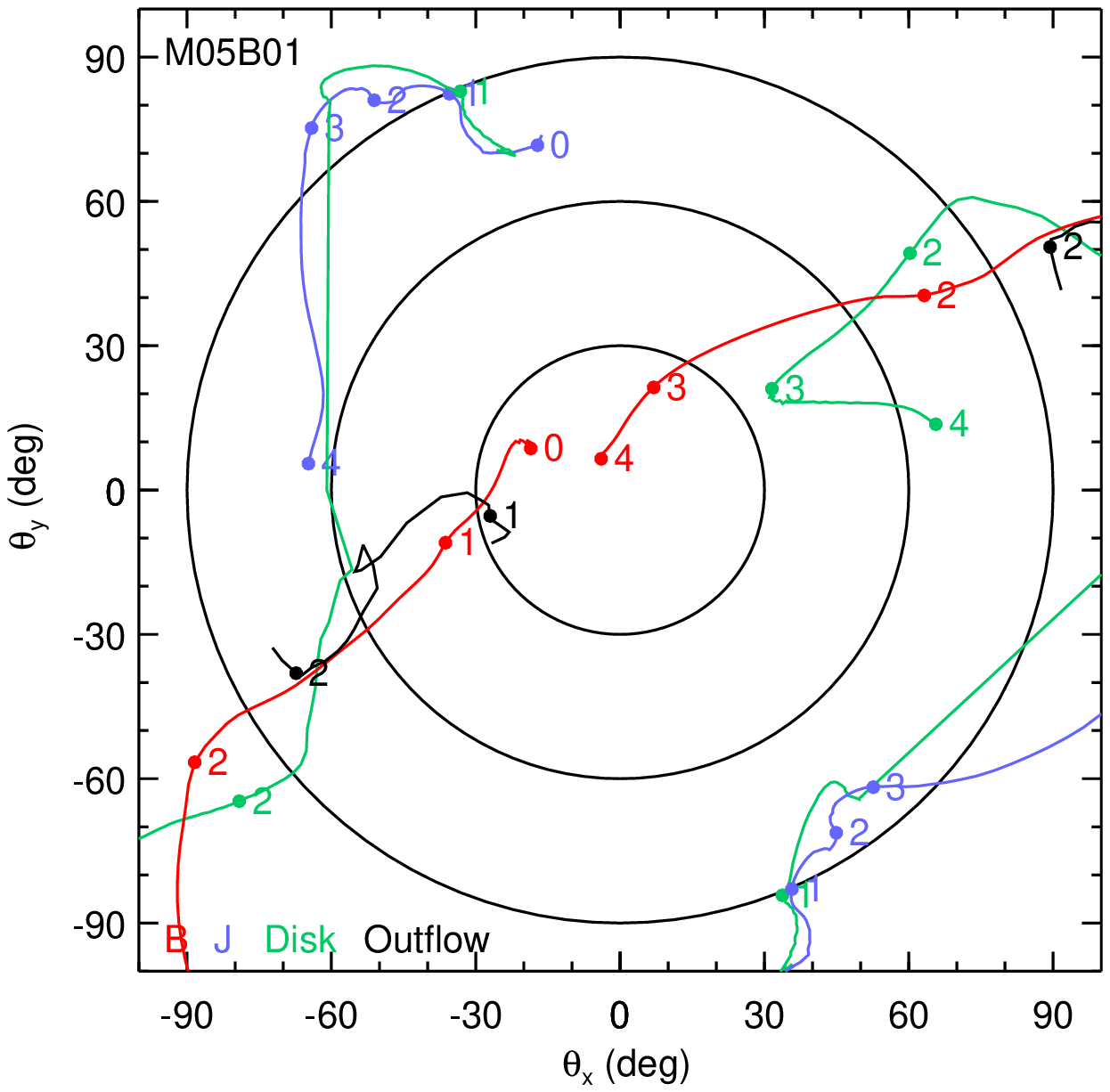}
  \plotone{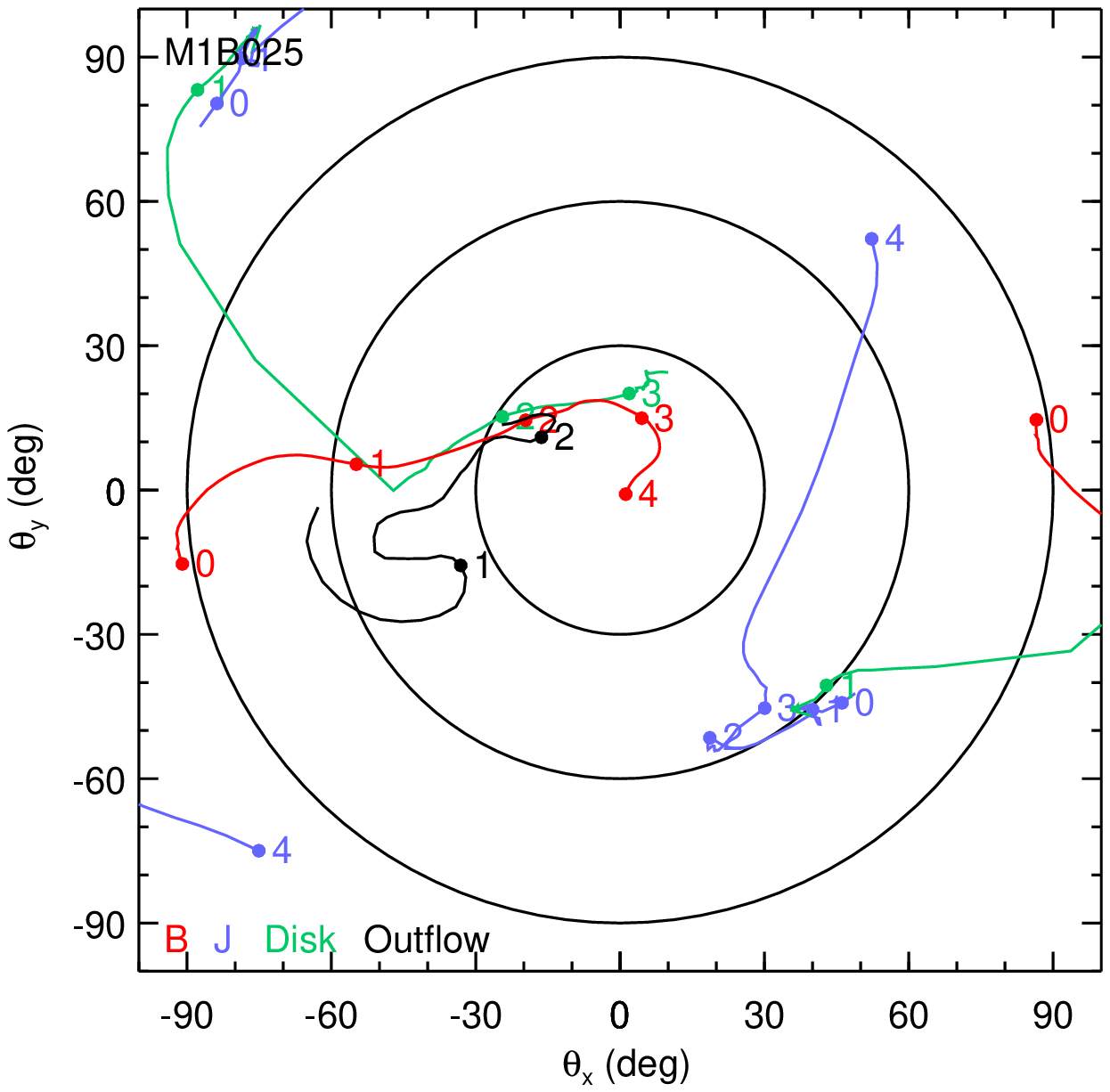}
\caption{Directions of the mean magnetic field (red lines), angular
  momentum (blue lines), minor axis of the density distribution (green
  lines), and outflow (black lines), as a function of the radius for models
  M05B01 and M1B025 at $t_p = 700$~yr.  The directions are
  plotted as $(\theta_x, \theta_y) = \arctan(V_{xy}/V_z) / V_{xy}
  (V_x, V_y)$ in the $\theta_x - \theta_y$ plane for a given vector
  $\bmath{V} = (V_x, V_y, V_z)$, where $V_{xy} = (V_x^2 +
  V_y^2)^{1/2}$.  The numbers associated with the filled circles indicate
  the distance from the sink particles as $\log (r/\mathrm{au})$.
  In order not to distinguish between parallel and antiparallel vectors , $\pm \bmath{V}$ are plotted for each vector $\bmath{V}$.
\label{directions.eps}
}
  \epsscale{0.5}
  \plotone{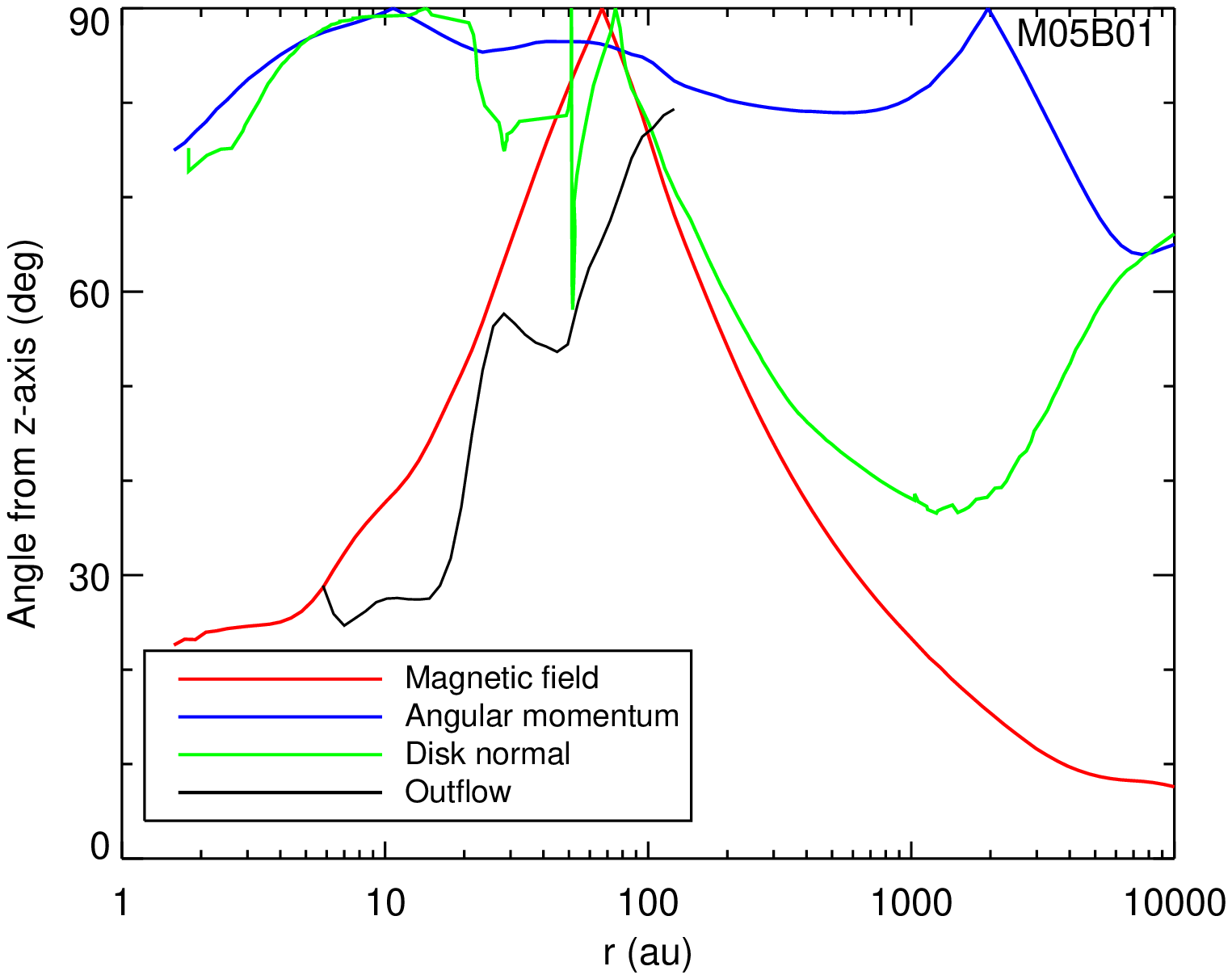}
  \plotone{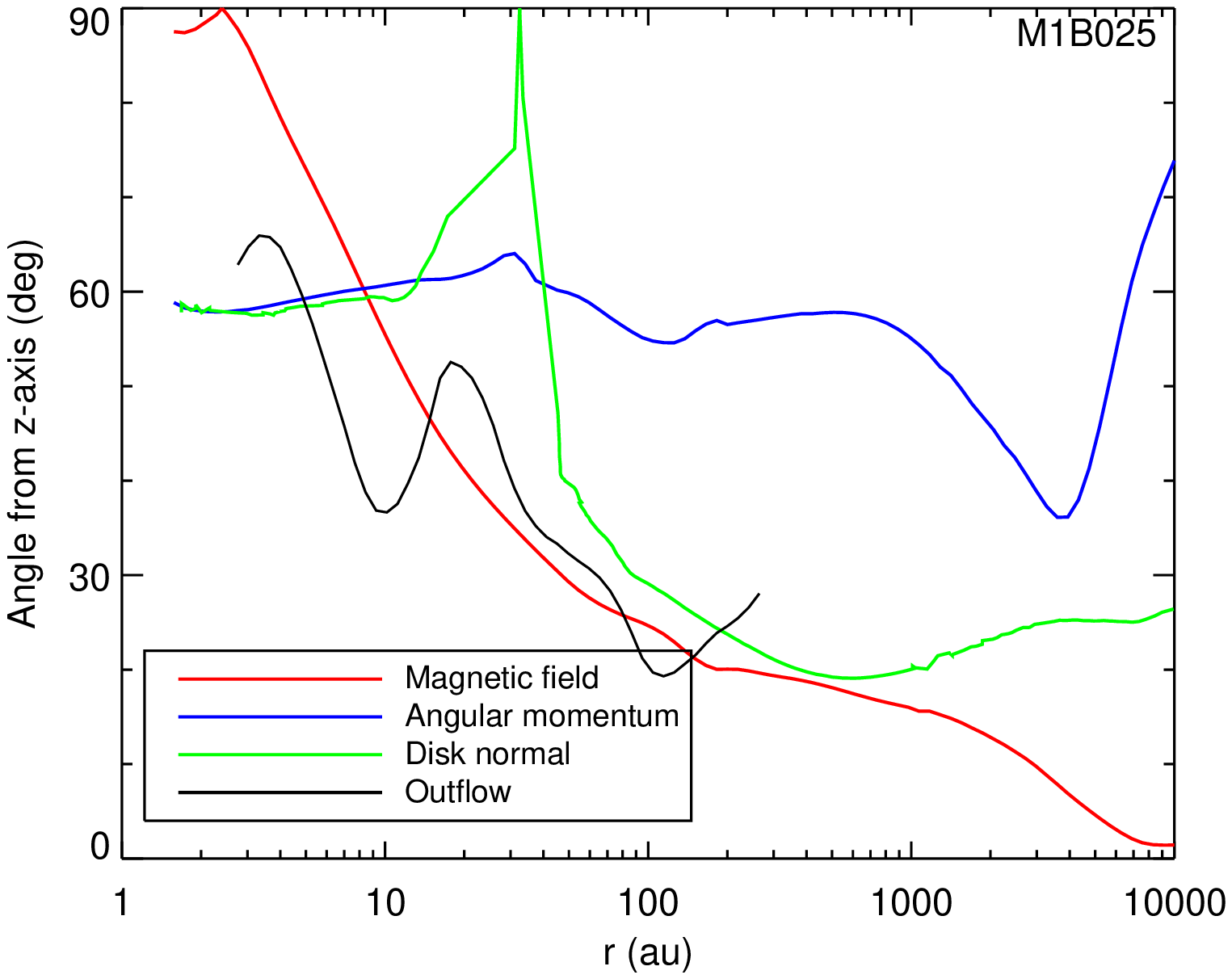}
  \caption{
    Orientations of the vector directions with respect to the $z$-axis
    as a function of the radius for 
    the mean magnetic field (red lines), angular
    momentum (blue lines), minor axis of the density distribution
    (green lines), and outflow (black lines) for models M05B01 and
    M1B025 at $t_p = 700$~yr.
    In order not to distinguish between parallel and antiparallel vectors , $\pm
    \bmath{V}$ are plotted for each vector $\bmath{V}$.
\label{directions_t.eps}
    }
\end{figure*}

Figures~\ref{directions.eps} and \ref{directions_t.eps} show the
directions of the magnetic field, the rotation axes, the disk-like
structures, and outflows for the representative models
M05B01 and M1B025. Measurement of these directions is described in
Appendix~\ref{sec:Measurement_of_the_directions_of_axes}.

For model M05B01, the mean magnetic field $\overline{\bmath{B}}$ on
the scale of $10^4$~au is antiparallel to that on the 1~au scale, indicating that the
magnetic field rotates up to $\sim 180\degr$ around the vector $\overline{\bmath{j}}$. 
The disk normal vector $\overline{\bmath{n}}$ is
associated with $\overline{\bmath{B}}$ on scales larger than 10~au,
indicating that the flattened envelope is perpendicular to the local
magnetic field.  On scales smaller than 10~au, 
the disk normal vector $\overline{\bmath{n}}$ is associated with the
mean angular momentum $\overline{\bmath{j}}$, indicating that the disk
is aligned with the rotation axis.  The outflow is accelerated on the
10~au scale, and it extends along the magnetic field, up to the 100~au
scale.  The outflow is therefore misaligned with the rotating disk, but it is 
aligned with the flattened envelope.

\begin{figure}
\epsscale{1.0}
\plotone{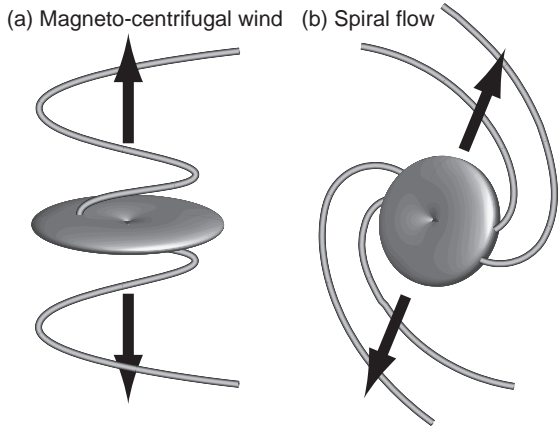}
\caption{Schematic diagram of two types of outflows: (a) magnetocentrifugal wind,
  and (b) spiral flow. The surfaces represent isodensity surfaces, and the
  tubes denote the magnetic field lines. The arrows indicate the
  direction of the outflow.
\label{flows.eps}
}
\end{figure}

The ejection mechanism for this outflow is different from the ordinal
magnetocentrifugal wind \citep{Blandford82,Pudritz86}; this outflow mechanism is 
a spiral flow (see Figure~\ref{flows.eps}b).
The model here demonstrates that the spiral flow reproduces a bipolar outflow. Similar spiral flows 
have been observed \citep[see Figure~21 of][]{Matsumoto11}.  
To confirm that the spiral flow mechanism continues to drive the outflow on the timescale of $10^4$~yr, further
long-term simulations are necessary \reviseb{\citep[e.g.,][]{Seifried12,Seifried13}}.

For model M1B025, the outflow is driven by the magnetocentrifugal wind on the
$\sim 100$~au scale (see Figure~\ref{flows.eps}a), and the flow direction is aligned with the local magnetic
field $\overline{\bmath{B}}$, as shown in Figures~\ref{directions.eps}
and \ref{directions_t.eps}.  On this scale, the flattened envelope is
also aligned with the magnetic field.
On the scale of $\lesssim 10$~au, the disk normal vector $\overline{\bmath{n}}$ is
aligned with the mean angular momentum $\overline{\bmath{j}}$,
indicating that the disk is aligned with the rotation axis.
The flow direction on this scale is
contaminated by the velocity associated with the formation of a cavity 
(see the right bottom panel of Figure~\ref{jets.eps}). 

The direction of the angular momentum vector depends on the radius of the cloud
core, as shown in Figure~\ref{directions.eps}.
This indicates that 
the disk accretes lumps of gas for which the angular momentum vectors have highly nonuniform directions.
The angular momentum of the infalling
gas greatly affects the evolution of the disk, as shown in \citet{Vorobyov15}.
As the disk further evolves, its orientation and size are both 
expected to change.

Observations of misalignment between outflows, magnetic
field, circumstellar disks, and flattened envelopes have been reported.
\citet{Hull13} showed that outflows are misaligned 
or randomly aligned relative to the  
magnetic field on the $\sim 1000$~au scale for Class 0 and Class I objects, and this is consistent with the results of the models considered here, which predicted that
the direction of the magnetic field
depends on the scale length (see, e.g., Figure~\ref{directions_t.eps}), though 
the outflow accelerates along the magnetic field on the $10-100$~au scale. 
If the outflow is aligned with the acceleration, 
it is expected to be misaligned with the magnetic field on the
$\sim 1000$~au scale.

Observations of the Class I source L1489 IRS have suggested that the
central Keplerian disk is inclined with respect to the flattened
infalling envelope \citep{Brinch07a,Brinch07b}.  This misalignment
between the central disk and the flattened envelope was
reproduced in all the models we considered, as shown in Figure~\ref{directions.eps}.
In each model, the disk normal vector $\overline{\bmath{n}}$ (the green line) drifts on the
$\theta_x - \theta_y$ plane, indicating that with the
flattened density structure, the inclination depends on the
radius.  Moreover, $\overline{\bmath{n}}$ is 
aligned with $\overline{\bmath{B}}$ when $r \gtrsim 100$~au,
suggesting that the inclination of the flattened envelope is guided by the
magnetic field. In other words, the flattened density structure of the
envelope is caused by the magnetic field.

\revise{
The Class I binary source in the Ophiuchus star-forming region, IRS~43, 
is a complex misaligned system.
The most curious misalignment in this object is that of the
circumbinary disk and the orbit of the binary.
According to \citet{Brinch16}, IRS~43 has a circumbinary
disk of which inclination is nearly edge-on, while the
orbit of the binary is close to being in the plane of the sky.
Such misalignment is possibly produced by a non-monotonic distribution
of the angular momentum on the scale of the cloud core as shown in 
Figure~\ref{directions.eps}.  On the cloud scale, the infalling gas
has misaligned angular momentum, leading the time-dependent accretion
of angular momenta onto the circumbinary disk and circumstellar disks.
}
Such a misaligned system cannot be reproduced
by axisymmetric models \citep[e.g.,][]{Machida08}.  
Even if weak turbulence is assumed, misaligned systems are reproduced.

\subsection{Cavity and arc-like structure}

The rim of the cavity has a higher column density than does the envelope, as shown in
Figure~\ref{colmndensity_M1B025_50au.eps}, and 
it may be observed as an arc-like structure.
\citet{Tokuda14} reported that the ALMA Cycle 0 observations reveal
an arc-like structure at the center of the high-density molecular cloud core
MC27 or L1521F. 
The arc-like structure was extended to a length of $\sim
1000$~au, and they proposed that it was caused
by a dynamical interaction between the dense gas condensation and
the envelope.  \citet{Matsumoto15b} performed 
hydrodynamical simulations to determine the origin of the arc-like
structure, and they demonstrated that 
gravitational torque due to the orbiting protostars produces arc-like
structures extending up to 1000~au. 

The typical length of the cavity is consistent with the
observations of MC27/L1521F.
The rim of the cavity obtained with the model M1B025 was extended
to $\sim 50$~au at $t_p = 1000$~yr. 
If we assume that the cavity continues to extend at a constant velocity of
$0.2\,\mathrm{km\,s}^{-1}$ (see Section~\ref{sec:Cavity_formation}), 
it takes $2\times 10^4$~yr for the rim to extended to 1000~au.
This timescale agrees with that of the arc structures
reproduced by \citet{Matsumoto15b}, and it is also
consistent with the timescale of the protostar ({\it Spizer} source)
in MC27/L1521F.
\revise{
  As shown in the literature
  \citep[e.g.,][]{Zhao11,Joos12,Krasnopolsky12,Machida14}, 
  the detailed structure of the cavity and the rim seems to be
  sensitive to the simulation settings.  
  Comparison between the models and the
  observations should be performed in terms of typical values, e.g., a
  typical length, as shown here.
  }

The rim of the cavity can account for the dense gas condensations in MC27/L1521F.
Our simulation results indicate that 
the number density of the rim is $\sim 10^{10}\,\mathrm{cm}^{-3}$ on
the 50~au scale.  
When the cavity is expanded up to 1000~au, the number
density of the rim is expected to be $2.5 \times 10^{7}\,\mathrm{cm}^{-3}$,
assuming that the density is distributed as $\rho \propto r^{-2}$.
The number densities of the dense gas condensations,
MMS-2 and MMS-3, are estimated to be  $10^{7}\,\mathrm{cm}^{-3}$ and
$10^{6-7}\,\mathrm{cm}^{-3}$, respectively \citep{Tokuda14}.
Therefore, \reviseb{MMS-2} and \reviseb{MMS-3} can be explained by the dense portions of
the rim rather than by the fragments. 

On the other hand, between the rim and the cavity, the column density differs by a factor of
$\sim 30$; the column densities are 
$\sim 5\times 10^{24}\,\mathrm{cm}^{-3}$ on the rim and 
$\sim 2\times 10^{23}\,\mathrm{cm}^{-3}$ in the cavity
(Figure~\ref{colmndensity_M1B025_50au.eps}).
Such a high contrast in the column density has not been seen in
observations of the ALMA Cycle~1 \citep[see Figure~3b of][]{Tokuda16}. %
Observations of the magnetic field of the cloud core will be 
of key importance in determining which model best accounts for the origin of
the arc-like structure.

\section{Summary}
\label{sec:summary}
Gravitational collapse of molecular cloud cores and the formation of
circumstellar disks and outflows around protostars were
investigated by performing AMR
simulations; the effects of both turbulence and the magnetic field were considered.
Ohmic dissipation was considered in the MHD simulations.
We allowed the system to evolve for $\sim 1000$~yr following the formation of 
  a protostar.
The main outcomes are summarized as follows.
\begin{enumerate}
\item In each of the magnetized models, the cloud core collapses to form a
  protostar surrounded by a circumstellar disk.
  The star-disk system is surrounded by an infalling envelope, and bipolar outflows are ejected.
  The nonmagnetized models produce massive circumstellar
  disks, one of which undergoes fragmentation at $\sim 10^4$~yr
  following the formation of a protostar.
\item The radius of the circumstellar disk depends on the initial strength of the
  magnetic field.  Models with a stronger magnetic field
  produce a circumstellar disk with a smaller radius. The mass of the
  disk shows a similar dependence on the magnetic field, where a
  stronger field produces a less massive disk.  The ratio of disk
  mass to stellar mass remains roughly constant at about $\sim 1 - 10$\%, depending on the strength of the magnetic
  field.  
\item The magnetized models reproduce the outflow, which can be
  classified into two types: a magnetocentrifugal wind and a spiral
  flow.  In the latter, the outflow is not aligned with the rotational
  axis of the disk.  In both the cases, the outflow and the
  flattened envelope are aligned with the magnetic field on that scale.
  In some models, the outflow is misaligned with the
  circumstellar disk.  Similarly, the
  flattened envelope may be misaligned with the circumstellar disk.
\item The internal distribution of angular momentum in the
  cloud cores is nonuniform. After long-term evolution, the disk
  accretes lumps of gas in which the direction of the angular
  momentum vectors is highly nonuniform; hence, the disk is expected to change its
  orientation and size.  This means that a planet formed during a later phase may have an orbital angular momentum that is highly misaligned with the angular momentum of the
  central star.
\item A strong magnetic field tends to produce a cavity
  in the infalling envelope; this is due to the strong magnetic
  pressure, and the gas accumulates on the rim.  Thus, the rim can account
  for the arc-like structure and dense gas condensation observed in the
  high-density molecular cloud core  MC27/L1521F, though it has not been verified by observation that there is a high contrast between the
  column density of the cavity and that of the rim.
\end{enumerate}

\acknowledgments 
We would like to thank K. Tokuda, T. Onishi, K. Tomida, R. Kawabe, and
N. Ohashi for fruitful discussions.
Numerical computations were carried out on the Cray XC30 and XT4 at
the Center for Computational Astrophysics, National Astronomical
Observatory of Japan, 
and 
the HITACHI HA8000 Clustre System (T2K-Todai) in the Information Technology Center, The University of Tokyo.
This research was supported by
JSPS KAKENHI Grant Numbers
16H02160,
15K05032, 
26400233,
26287030,
25400232, 
24244017,
23540270, and
23244027.


\begin{appendix}
  \section{Measurement of the Disk Radius}
  \label{sec:Measurements_of_the_Disk_Radii}
The disk radius for each model was estimated using the density and velocity
distributions, as follows.
The volume of the disk $V_d$ was defined by two criteria:
$\rho \ge \rho_\mathrm{cr}$ and 
$(v_\varphi^2 + v_\theta^2)^{1/2}/(v_r^2 + c_s^2)^{1/2} \ge 3$.
The former indicates that the disk has a density higher than the
critical density of the equation of state.
The latter indicates that the velocity of rotation is greater than the
radial velocity.  In spherical coordinates, the velocity is
$(v_r, v_\theta, v_\varphi)$, 
which was calculated by a transformation from Cartesian coordinates with the origin set 
at the location of the sink particle.  Because the disk orientation is not aligned to any of the 
coordinate axes, the tangential velocity of $(v_\varphi^2 + v_\theta^2)^{1/2}$
was adopted as the rotational velocity.
The value of three on the right-hand side of the second criterion was determined empirically. 

The disk radius was obtained from the inertia tensor of the volume, $V_d$.
The inertia tensor is calculated as
\begin{equation}
  I =
  \left(
  \begin{array}{ccc}
    I_{xx} & I_{xy} & I_{xz}\\
    I_{yx} & I_{yy} & I_{yz}\\
    I_{zx} & I_{zy} & I_{zz}\\
  \end{array}
  \right)\;,
\label{eq:inertia_matrix}
\end{equation}
and each element of the matrix is calculated by a moment of the
coordinates, e.g.,
\begin{equation}
I_{xy} =
\frac{\int_{V_d} \left( x - x_p \right)\left( y - y_p \right)dV}{\int_{V_d}dV}
\;,
\label{eq:appendix_ixx}
\end{equation}
where $(x, y, z)$ are the coordinates of a cell, and 
$(x_p, y_p, z_p)$ is the position vector for a sink particle.  The
volume integrals in equation~(\ref{eq:appendix_ixx}) were performed by
summing over the cells within the 
volume, $V_d$.
The matrix $I$ yields three eigenvalues, 
$\lambda_1 > \lambda_2 > \lambda_3$, 
and the square root of each of the eigenvalues corresponds to the 
length of a principal axis;
$\lambda_1^{1/2}$, $\lambda_2^{1/2}$, and $\lambda_3^{1/2}$
correspond to the semi-major axis, 
the semi-minor axis, and the thickness of the disk, respectively.
The semi-minor axis $\lambda_2^{1/2}$ was adopted as the disk radius
so that this method could be applied to a highly elongated disk, such as the disk of
model M1B025 (the lower right panel of Figure~\ref{25au.eps}).
The disk radius is defined as
\begin{equation}
R_d = 2 \lambda_2^{1/2}\;,
\end{equation}
where the factor of two comes from the inertial tensor for a uniform thin disk
with a radius of $R_d$, i.e., $I_{xx} + I_{yy} = (1/2) R_d^2$.

We also calculated the mass of the disks using
\begin{equation}
M_d = \int_{V_d} \rho dV\;,
\end{equation}
where  $V_d$ is the disk volume.
\section{Measurement of the direction of the axes}
\label{sec:Measurement_of_the_directions_of_axes}
In Figures~\ref{directions.eps} and \ref{directions_t.eps}, the direction of the magnetic field, angular momentum, disk,
and outflow are each shown as a function of the radius.  These axes were measured as follows.

In order to define the direction of the magnetic field and angular
momentum, we measured the mean magnetic field and the angular momentum: 
\begin{align}
\overline{\bmath{B}}(r) &= \frac{1}{V_s(r)} \int_{V_s(r)} \bmath{B}(\bmath{r}) \,dV,\\
\overline{\bmath{j}}(r) &= \frac{1}{M(r)} \int_{V_s(r)} \rho(\bmath{r})\, (\bmath{r}-\bmath{r}_p)
\times (\bmath{v}(\bmath{r}) - \bmath{v}_p) \,dV. 
\end{align}
where 
\begin{equation}
M(r) = \int_{V_s(r)} \rho \,dV,
\end{equation}
The volume $V_s(r)$ is that of a sphere with radius $r$, and the
center coincides with the position of the sink particle:
\begin{equation}
V_s(r) = \{ \bmath{r} \in \mathbb{R}^3 \mid
|\bmath{r} - \bmath{r}_p| \le r \}.
\end{equation}
The vectors $\bmath{r}_p$ and $\bmath{v}_p$, respectively, denote the position and velocity of the sink particle.

The orientation of the disk was calculated using the eigenvector of the
inertia tensor, which is similar to $I$ in equation (\ref{eq:inertia_matrix}).
The elements of the inertia tensor are obtained as follows.
\begin{equation}
I_{xy} =
\frac{\int_{\rho\ge\rho_d} \rho \left( x - x_p \right)\left( y - y_p
  \right)dV}{\int_{\rho\ge\rho_d} \rho dV}
\;,
\label{eq:appendix_iixx}
\end{equation}
where the integration is performed inside the region in which $\rho \ge
\rho_d$, for a given threshold $\rho_d$.  The eigenvector associated with the
smallest eigenvalue represents a normal vector for the flattened
disk-like structure.  The radius of the disk-like structure is defined as the maximum
extent of the region in which $\rho \ge \rho_d$, as measured from the position of the sink
particle.  Thus, we obtain the direction of the disk normal vector $\overline{\bmath{n}}$ as a function of
the radius $r$.  Note that the disk measured here corresponds to the
circumstellar disk for a small radius, e.g., $r \la 10$~au,  and to the flattened infalling envelope for a large radius.

The direction of the outflow was obtained from the flow of the gas inside
the outflow region.  Because the outflow is bipolar and the gas flows
roughly parallel to the magnetic field, the direction of the outflow is
calculated as follows.
\begin{equation}
\overline{\bmath{v}}_\mathrm{of}(r) = \frac{1}{V_\mathrm{of}(r)}\int_{V_\mathrm{of}(r)} \bmath{v} \, \mathrm{sign}(\bmath{v}\cdot\overline{\bmath{B}})\,dV,
\end{equation}
where $V_\mathrm{of}(r)$ denotes the region of the outflow, defined as
\begin{equation}
V_\mathrm{of}(r) = \{\bmath{r} \in V_s(r) \mid v_r(\bmath{r}) \ge 2 c_s\}.
\end{equation}

\end{appendix}


\begin{thebibliography}{}
\bibitem[Aso et al.(2015)]{Aso15} Aso, Y., Ohashi, N., Saigo, K., et al.\ 2015, \apj, 812, 27 
\bibitem[Belloche et al.(2006)]{Belloche06} Belloche, A., Parise, B., van der Tak, F.~F.~S., et al.\ 2006, \aap, 454, L51 
\bibitem[Blandford \& Payne(1982)]{Blandford82} Blandford, R.~D., \& Payne, D.~G.\ 1982, \mnras, 199, 883 
\bibitem[Bonnor(1956)]{Bonnor1956} Bonnor, W. B. 1956, \mnras, 116,351
\bibitem[Brinch et al.(2007a)]{Brinch07a} Brinch, C., Crapsi, A., Hogerheijde, M.~R., \& J{\o}rgensen, J.~K.\ 2007a, \aap, 461, 1037 
\bibitem[Brinch et al.(2007b)]{Brinch07b} Brinch, C., Crapsi, A., J{\o}rgensen, J.~K., Hogerheijde, M.~R., \& Hill, T.\ 2007b, \aap, 475, 915 
\bibitem[Brinch et al.(2016)]{Brinch16} Brinch, C., J{\o}rgensen, J.~K., Hogerheijde, M.~R., Nelson, R.~P., \& Gressel, O.\ 2016, \apjl, 830, L16 
\bibitem[Burkert \& Bodenheimer(2000)]{Burkert00} Burkert, A., \& Bodenheimer, P.\ 2000, \apj, 543, 822 
\bibitem[Chandrasekhar(1939)]{Chandrasekhar39} Chandrasekhar, S.\ 1939, An Introduction to the Study of Stellar Structure, (Chicago, Ill., Univ. Chicago press)
\bibitem[Ciardi \& Hennebelle(2010)]{Ciardi10} Ciardi, A., \& Hennebelle, P.\ 2010, \mnras, 409, L39 
\bibitem[Crutcher(1999)]{Crutcher99} Crutcher, R.~M.\ 1999, \apj, 520, 706 
\bibitem[Dubinski et al.(1995)]{Dubinski95} Dubinski, J., Narayan, R., \& Phillips, T.~G.\ 1995, \apj, 448, 226 
\bibitem[Ebert(1955)]{Ebert1955} Ebert, R. 1955, Z. Astrophys., 37, 222
\bibitem[Federrath et al.(2011)]{Federrath11} Federrath, C., Sur, S., Schleicher, D.~R.~G., Banerjee, R., \& Klessen, R.~S.\ 2011, \apj, 731, 62 
\bibitem[Fukuda \& Hanawa(1999)]{Fukuda99} Fukuda, N., \& Hanawa, T.\ 1999, \apj, 517, 226 
\bibitem[Hennebelle \& Ciardi(2009)]{Hennebelle09} Hennebelle, P., \& Ciardi, A.\ 2009, \aap, 506, L29 
\bibitem[Hogerheijde(2001)]{Hogerheijde01} Hogerheijde, M.~R.\ 2001, \apj, 553, 618 
\bibitem[Hull et al.(2013)]{Hull13} Hull, C.~L.~H., Plambeck, R.~L., Bolatto, A.~D., et al.\ 2013, \apj, 768, 159 
\bibitem[Joos et al.(2012)]{Joos12} Joos, M., Hennebelle, P., \& Ciardi, A.\ 2012, \aap, 543, A128 
\bibitem[J{\o}rgensen et al.(2009)]{Jorgensen09} J{\o}rgensen, J.~K., van Dishoeck, E.~F., Visser, R., et al.\ 2009, \aap, 507, 861 
\bibitem[Krasnopolsky et al.(2012)]{Krasnopolsky12} Krasnopolsky, R., Li, Z.-Y., Shang, H., \& Zhao, B.\ 2012, \apj, 757, 77 
\bibitem[Larson(1981)]{Larson81} Larson, R.~B.\ 1981, \mnras, 194, 809 
\bibitem[Larson(1985)]{Larson85} Larson, R.~B.\ 1985, \mnras, 214, 379 
\bibitem[Li \& McKee(1996)]{Li96} Li, Z.-Y., \& McKee, C.~F.\ 1996, \apj, 464, 373 
\bibitem[Li et al.(2013)]{Li13} Li, Z.-Y., Krasnopolsky, R., \& Shang, H.\ 2013, \apj, 774, 82 
\bibitem[Machida \& Hosokawa(2013)]{Machida13} Machida, M.~N., \& Hosokawa, T.\ 2013, \mnras, 431, 1719 
\bibitem[Machida et al.(2007)]{Machida07} Machida, M.~N., Inutsuka, S.-i., \& Matsumoto, T.\ 2007, \apj, 670, 1198 
\bibitem[Machida et al.(2008)]{Machida08} Machida, M.~N., Inutsuka, S.-i., \& Matsumoto, T.\ 2008, \apj, 676, 1088-1108 
\bibitem[Machida et al.(2011)]{Machida11} Machida, M.~N., Inutsuka, S.-I., \& Matsumoto, T.\ 2011, \pasj, 63, 555 
\bibitem[Machida et al.(2014)]{Machida14} Machida, M.~N., Inutsuka, S.-i., \& Matsumoto, T.\ 2014, \mnras, 438, 2278 
\bibitem[Masunaga, Miyama, \& Inutsuka(1998)]{Masunaga98} Masunaga, H., Miyama, S.~M., \& Inutsuka, S.\ 1998, \apj, 495, 346. 
\bibitem[Matsumoto \& Hanawa(2011)]{Matsumoto11} Matsumoto, T., \& Hanawa, T.\ 2011, \apj, 728, 47 
\bibitem[Matsumoto \& Tomisaka(2004)]{Matsumoto04} Matsumoto, T., \& Tomisaka, K.\ 2004, \apj, 616, 266 
\bibitem[Matsumoto et al.(2015a)]{Matsumoto15a} Matsumoto, T., Dobashi, K., \& Shimoikura, T.\ 2015a, \apj, 801, 77 
\bibitem[Matsumoto et al.(2015b)]{Matsumoto15b} Matsumoto, T., Onishi, T., Tokuda, K., \& Inutsuka, S.-i.\ 2015b, \mnras, 449, L123 
\bibitem[Matsumoto(2007)]{Matsumoto07} Matsumoto, T.\ 2007, \pasj, 59, 905 
\bibitem[Matsumoto(2011)]{Matsumoto11b} Matsumoto, T.\ 2011, \pasj, 63, 317 
\bibitem[Mellon \& Li(2008)]{Mellon08} Mellon, R.~R., \& Li, Z.-Y.\ 2008, \apj, 681, 1356-1376 
\bibitem[Miyoshi \& Kusano(2005)]{Miyoshi05} Miyoshi, T., \& Kusano, K.\ 2005, Journal of Computational Physics, 208, 315 
\bibitem[Murillo et al.(2013)]{Murillo13} Murillo, N.~M., Lai, S.-P., Bruderer, S., Harsono, D., \& van Dishoeck, E.~F.\ 2013, \aap, 560, A103 
\bibitem[Nakano \& Nakamura(1978)]{Nakano78} Nakano, T., \& Nakamura, T.\ 1978, \pasj, 30, 671 
\bibitem[Ohashi et al.(2014)]{Ohashi14} Ohashi, N., Saigo, K., Aso, Y., et al.\ 2014, \apj, 796, 131 
\bibitem[Onishi et al.(1998)]{Onishi98} Onishi, T., Mizuno, A., Kawamura, A., Ogawa, H., \& Fukui, Y.\ 1998, \apj, 502, 296 
\bibitem[Pineda et al.(2011)]{Pineda11} Pineda, J.~E., Arce, H.~G., Schnee, S., et al.\ 2011, \apj, 743, 201 
\bibitem[Pudritz \& Norman(1986)]{Pudritz86} Pudritz, R.~E., \& Norman, C.~A.\ 1986, \apj, 301, 571 
\bibitem[P{\'e}rez et al.(2016)]{Perez16} P{\'e}rez, L.~M., Carpenter, J.~M., Andrew, S.~M., et al.\ 2016, Science, 353, 1519
\reviseb{\bibitem[Seifried et al.(2012)]{Seifried12} Seifried, D., Banerjee, R., Pudritz, R.~E., \& Klessen, R.~S.\ 2012, \mnras, 423, L40}
\reviseb{\bibitem[Seifried et al.(2013)]{Seifried13} Seifried, D., Banerjee, R., Pudritz, R.~E., \& Klessen, R.~S.\ 2013, \mnras, 432, 3320}
\bibitem[Tassis \& Mouschovias(2005)]{Tassis05} Tassis, K., \& Mouschovias, T.~C.\ 2005, \apj, 618, 783 
\bibitem[Tobin et al.(2012)]{Tobin12} Tobin, J.~J., Hartmann, L., Chiang, H.-F., et al.\ 2012, \nat, 492, 83 
\bibitem[Tokuda et al.(2014)]{Tokuda14} Tokuda, K., Onishi, T., Saigo, K., et al.\ 2014, \apjl, 789, L4 
\bibitem[Tokuda et al.(2016)]{Tokuda16} Tokuda, K., Onishi, T., Matsumoto, T., et al.\ 2016, \apj, 826, 26 
\bibitem[Tomida et al.(2017)]{Tomida17} Tomida, K., Machida, M.~N., Hosokawa, T., Sakurai, Y., \& Lin, C.~H.\ 2017, \apjl, 835, L11 
\bibitem[Tomisaka et al.(1988)]{Tomisaka88} Tomisaka, K., Ikeuchi, S., \& Nakamura, T.\ 1988, \apj, 335, 239 
\bibitem[Toomre(1964)]{Toomre64} Toomre, A.\ 1964, \apj, 139, 1217 
\bibitem[Truelove et al.(1997)]{Truelove97} Truelove, J.~K., Klein, R.~I., McKee, C.~F., Holliman, J.~H., II, Howell, L.~H., \& Greenough, J.~A.\ 1997, \apjl, 489, L179 
\bibitem[Vorobyov et al.(2015)]{Vorobyov15} Vorobyov, E.~I., Lin, D.~N.~C., \& Guedel, M.\ 2015, \aap, 573, A5 
\bibitem[Yen et al.(2014)]{Yen14} Yen, H.-W., Takakuwa, S., Ohashi, N., et al.\ 2014, \apj, 793, 1 
\bibitem[Zhao et al.(2011)]{Zhao11} Zhao, B., Li, Z.-Y., Nakamura, F., Krasnopolsky, R., \& Shang, H.\ 2011, \apj, 742, 10 
\bibitem[Zuckerman \& Evans(1974)]{Zuckerman74} Zuckerman, B., \& Evans, N.~J., II 1974, \apjl, 192, L149 

\end{thebibliography}
\end{document}